\documentclass[12pt]{article}
\usepackage{scicite}
\usepackage{array}
\usepackage[a4paper, text={16.5cm, 25.2cm}, centering]{geometry}
\usepackage[labelfont=bf]{caption}
\usepackage{varwidth}
\DeclareCaptionLabelSeparator{pipe}{ | }
\usepackage{amsmath}
\usepackage{booktabs,subcaption,amsfonts,dcolumn}
\usepackage{makecell}
\usepackage{url}
\usepackage{diagbox}
\usepackage{mathptmx}
\usepackage{anyfontsize}
\usepackage{t1enc}
\usepackage{diagbox}
\usepackage{booktabs}
\usepackage{blindtext}
\usepackage{hyperref}
\usepackage{float}
\usepackage{tabularx}
\usepackage[linesnumbered, ruled, vlined]{algorithm2e}
\usepackage[group-separator={,},group-minimum-digits={3}]{siunitx}

\SetAlgorithmName{Supplementary Algorithm}{Supplementary Algorithm}{List of Supplementary Algorithms}

\newcolumntype{x}[1]{>{\centering\arraybackslash\hspace{0pt}}p{#1}}

\usepackage{url}
\usepackage[colorinlistoftodos]{todonotes}
\usepackage{times}
\usepackage{gensymb}
\usepackage{lineno}

\usepackage{graphicx}
\usepackage{booktabs}
\usepackage{multirow}
\usepackage{threeparttable}
\usepackage{booktabs}    % for professional horizontal lines
\usepackage{enumitem}

\usepackage[utf8]{inputenc}

\DeclareMathOperator*{\argmaxA}{arg\,max} % Jan Hlavacek
\newcommand{\red}[1]{\textcolor{black}{#1}}

\newcolumntype{M}[1]{>{\centering\arraybackslash}m{#1}}

\newcommand{\pttscgwrist}{PTT_{SCG \rightarrow wrist}}
\newcommand{\pttscgneck}{PTT_{SCG \rightarrow neck}}
\newcommand{\pttbcgwrist}{PTT_{BCG \rightarrow wrist}}
\newcommand{\pttbcgneck}{PTT_{BCG \rightarrow neck}}

\newcommand{\sysname}{\textit{PolyPulse}}

\newcommand{\squishlist}{\begin{itemize}[itemsep=1pt,parsep=2pt,topsep=3pt,partopsep=0pt,leftmargin=0em, itemindent=1em,labelwidth=1em,labelsep=0.5em]}
\newcommand{\squishend}{\end{itemize}}

\newenvironment{sciabstract}{%
\begin{quote} \bf}
{\end{quote}}

\title{Measuring multi-site pulse transit time with an\\ AI-enabled mmWave radar}

\date{}

\author{Jiangyifei Zhu$^{1\dag}$, Kuang Yuan$^{1\dag}$, Akarsh Prabhakara$^{2}$, Yunzhi Li$^{1}$, Gongwei Wang$^{1}$,\\ Kelly Michaelsen$^{3}$, Justin Chan$^{1\ast}$, Swarun Kumar$^{1\ast}$\\
\small
{$^{1}$Department of Electrical and Computer Engineering, Carnegie Mellon University, Pittsburgh, PA, USA}\\
\small
{$^{2}$Department of Computer Sciences, University of Wisconsin-Madison, Madison, WI, USA}\\
\small
{$^{3}$Department of Anesthesiology \& Pain Medicine, University of Washington, Seattle, WA, USA}\\
\small
{$^\dag$Equal contribution first authors}\\
\small
{$^\ast$Corresponding authors: justinchan@cmu.edu, swarun@cmu.edu}
}

\begin{document}

% \linenumbers
\baselineskip24pt

\maketitle

\begin{sciabstract}
\noindent
Pulse Transit Time (PTT) is a measure of arterial stiffness and a physiological marker associated with cardiovascular function, with an inverse relationship to diastolic blood pressure (DBP). We present the first AI-enabled mmWave system for contactless \textit{multi-site} PTT measurement using a single radar. By leveraging radar beamforming and deep learning algorithms our system simultaneously measures PTT and estimates diastolic blood pressure at multiple sites. The system was evaluated across three physiological pathways -- heart-to-radial artery, heart-to-carotid artery, and mastoid area-to-radial artery -- achieving correlation coefficients of 0.73--0.89 compared to contact-based reference sensors for measuring PTT. Furthermore, the system demonstrated correlation coefficients of 0.90--0.92 for estimating DBP, and achieved a mean error of -1.00--0.62~mmHg and standard deviation of 4.97--5.70~mmHg, meeting the FDA's AAMI guidelines for non-invasive blood pressure monitors. These results suggest that our proposed system has the potential to provide a non-invasive measure of cardiovascular health across multiple regions of the body.
\end{sciabstract}

\vskip 0.4in
\section*{Introduction}
Frequent monitoring of cardiovascular health can facilitate timely interventions including lifestyle changes such as exercise and diet, as well as pharmacologic changes like antihypertensive drugs~\cite{boutouyrie2011pharmacological}. Increased arterial stiffness is associated with a higher risk of cardiovascular conditions including hypertension~\cite{safar2018arterial}, stroke~\cite{mattace2006arterial}, coronary artery disease~\cite{mattace2006arterial}, sleep apnea~\cite{katz_pulse_2003,pitson_value_1998}, and Alzheimer's disease~\cite{hughes2015review}, and can be quantified using \textit{pulse transit time (PTT)}, the time it takes for a pulse wave to travel between a proximal and distal site within an artery~\cite{smith1999pulse}. In addition, pulse transit time is {indirectly related to diastolic blood pressure (DBP)} and has been explored as the basis for cuff-less blood pressure monitoring~\cite{mukkamala2015toward,wang2018seismo,geddes1981pulse,peter2014review,foo2009clinical}. The ability to measure PTT contactlessly at multiple sites along the body has the potential to support pre-screening of cardiovascular diseases and long-term tracking of disease progression, particularly for high-risk populations such as older adults and individuals with chronic conditions~\cite{bahache2020inclusive}. This is because PTT varies across different body sites and site-specific difference are influenced by underlying physiologic stress mechanisms~\cite{manoj_measurement_2025}. This suggests that multiple PTT measurements may provide a more comprehensive assessment of cardiovascular hemodynamics than a single measurement~\cite{di2021multi}.

Arterial stiffness can be measured using contact-based methods, including pulse tonometry for arterial pressure assessment~\cite{kemmotsu_arterial_1991}, photoplethysmography (PPG) for detecting light absorption changes~\cite{allen2002age}, inertial measurement units (IMU) for capturing pulse motion~\cite{wang_quantitative_2020}, and electrodes or wearable sensors~\cite{holz2017glabella,carek2017seismowatch,wang2018seismo,carek2018naptics,ganti2020wearable,winokur2012wearable,yousefian2019potential,salvi2004validation}, which may cause skin irritation and discomfort. Contactless measurement systems have leveraged cameras~\cite{niu2023full,murakami2015non,huang2024camera,patil2019camera,block2020conventional} to measure PPG, and laser-doppler vibrometry~\cite{beeckman_enhancing_2023} to measure the surface displacement at multiple arterial sites. However, such optical-based systems may not work well in low-lighting conditions or when objects like clothing block the device's view. More recently, radar-based approaches have been proposed, leveraging millimeter-wave (mmWave)~\cite{antolinos2023pulse,singh2023remote,geng2023contactless,liang2023airbp,johnson2019arterial}, or microwave~\cite{yoshioka2020analysis}, to measure the seismocardiogram (SCG) at the chest and ballistocardiogram (BCG) at the head. However, existing radar-based systems either require multiple sensor units to measure multiple arterial points~\cite{antolinos2023pulse,singh2023remote}, which can be challenging to deploy and precisely position, or are limited to measuring a single PTT pair~\cite{liang2023airbp,hu_contactless_2024,cao_hbp-fi_2024}. 

Here, we present {\sysname}, a proof-of-concept AI-enabled mmWave radar system that contactlessly measures pulse transit time and estimates DBP along three key physiological pathways along the upper body. Our system transmits chirp signals toward a seated individual and performs fine-grained beamforming on the reflected signal to capture the subtle bodily movements caused by cardiac activity. The design leverages a neural network model to process the reflection waveforms from the apex of the heart, the mastoid area at the head, the radial artery at the wrist, and the carotid artery at the neck. Using these signals, the model jointly estimates pulse transit times and diastolic blood pressure for physiological pathways that originate from the heart and head and extend to distal points at the wrist or neck. Our wireless radar system is compact and measures 16.5 $\times$ 14~cm, which can fit within the form factor of smart home devices. Given the increasing prevalence of mmWave radars on consumer devices, our proposed multi-site PTT system can make remote monitoring of cardiovascular health at home more comprehensive and accessible. 

\section*{Results}
\subsection*{Concept and prototype}

\newpage
\thispagestyle{empty}
\begin{figure}[H]
\centering
{\includegraphics[width=0.75\textwidth]{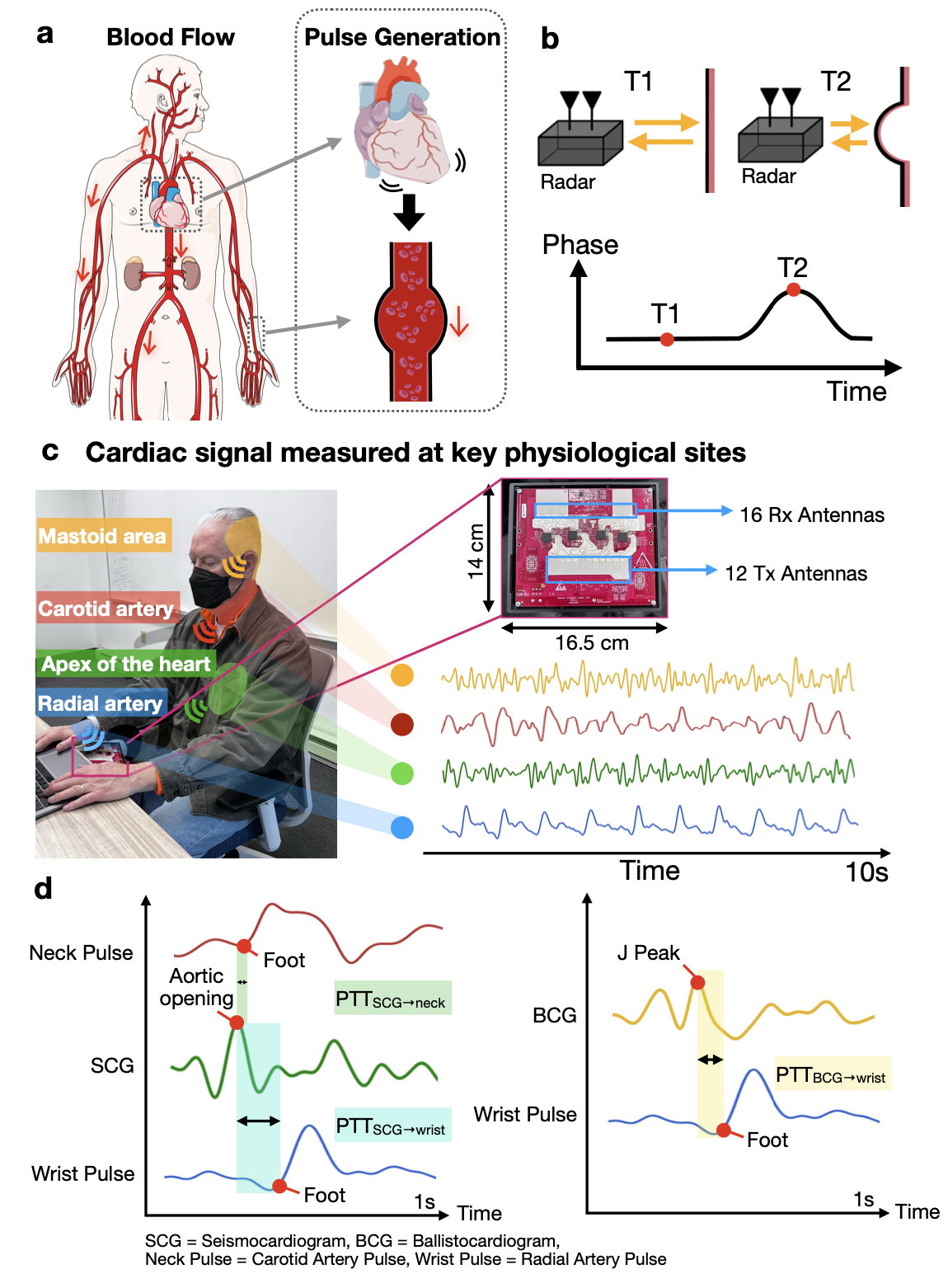}}
\vspace{-1em}
\caption{{\bf Overview of the mmWave radar system for multi-site pulse transit time (PTT) measurement.} {\bf a,} Rhythmic contractions of the heart generate pulse waves that propagate across the arterial system. (Image source: https://smart.servier.com/category/anatomy-and-the-human-body/cardiovascular-system) {\bf b,} The mmWave radar detects minute surface displacements by measuring phase variations in reflected signals over time. {\bf c,} The mmWave radar's beamforming algorithm targets four key physiological sites to measure PTT: the apex of the heart (green) and the mastoid area (yellow) as proximal reference points, and the radial artery (blue) and carotid artery (red) as distal arterial sites. The waveforms represent the phase signals from these sites, captured by the mmWave radar positioned at a fixed distance beneath the subject. {\bf d,} Our system estimates pulse transit time across three pairs: $\pttscgwrist, \pttscgneck, \pttbcgwrist$. This is achieved by first identifying key {cardiac waveforms features}: the aortic opening in the SCG signal, the J peak in the BCG signal, and the foot of the carotid and radial arterial pulse waveform. PTT is then computed as the time difference between these signal features at the proximal reference points and those at distal arterial sites.}
  \label{fig:overview}
\end{figure}

As the heart beats, it generates pulse waves that propagate through  arterial pathways in the body. This pulsatile flow causes rhythmic expansion and contraction of the arterial walls, which in turn results in minute superficial displacements of the skin (Fig.~\ref{fig:overview}a)~\cite{sharma_acoustic_2019}. {\sysname} transmits FMCW signals from 77--81~GHz towards a seated individual and analyzes both the phase and magnitude of the reflected signals at key physiological sites where a cardiac signal can be measured. The phase variations of the reflections correspond to superficial displacement of the skin at distal sites (\autoref{fig:overview}b, c), while the magnitude variations correspond to changes in signal absorption due to blood flow that occur throughout each cardiac cycle. Our system measures cardiac signals by capturing reflections from two proximal reference points and two distal arterial points. For the proximal reference point, our radar captures reflections from the apex of the heart at the chest to obtain the SCG signal, and the mastoid area at the head to obtain the BCG signal respectively. For the distal reference points, we measure reflections from the right radial artery at the wrist and the right carotid artery at the neck. To measure PTT we use the aortic valve opening (AO) in the SCG waveform and the J-peak in the BCG waveform as the starting time reference points, while the foot of the pulse waveforms at the distal sites serves as the ending reference (\autoref{fig:overview}d). By taking the time difference between pair of these starting and ending reference points, our radar system is able to measure PTT across three different pairs: heart-to-wrist ($\pttscgwrist$), heart-to-neck ($\pttscgneck$) and head-to-wrist ($\pttbcgwrist$). After measuring perturbations in PTT when the subject is at rest and following physical exercise, our system is able to derive an estimate of DBP along these pathways using a subject-specific calibration model~\cite{wibmer_pulse_2014}.

\begin{figure}[H]
\centering
{\includegraphics[width=\textwidth]{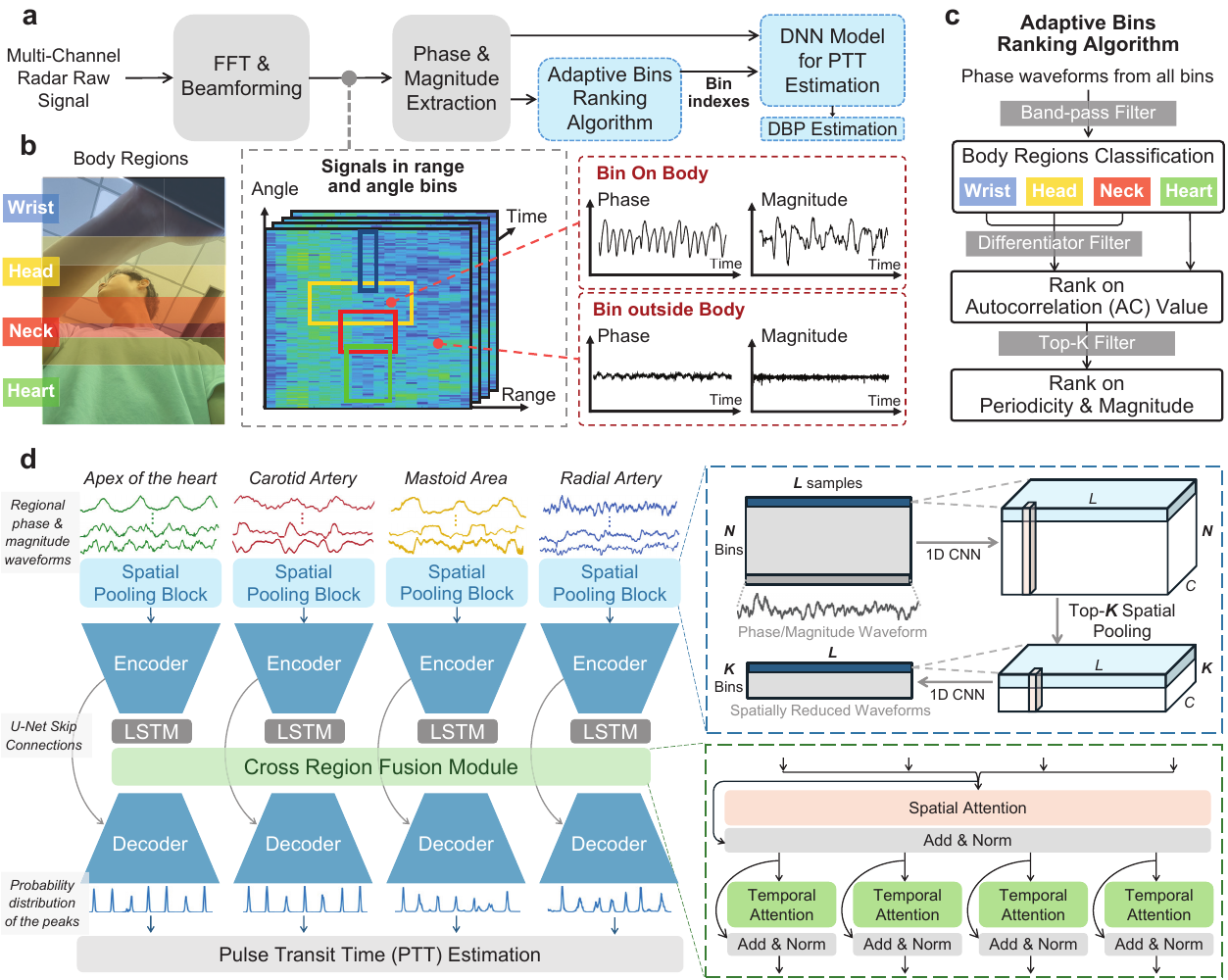}}
  \caption{{\bf Signal processing and deep neural network pipeline to estimate PTT from mmWave radar reflections along multiple physiological pathways.} {\bf a,} The system spatially decomposes the radar input channels into radar range and angular bins using FFT and beamforming. For each bin, phase and magnitude information are extracted. An adaptive bin ranking algorithm is applied to identify the bins with the strongest cardiac signal features, followed by a DNN model for PTT estimation. {\bf b,} Beamformed radar signal power is visualized across azimuth, range, and time axes, regions associated with higher signal power are brighter. Key physiological areas of interest, including the heart, neck, and wrist, are marked with bounding boxes that correspond to pre-defined search areas for a cardiac signal. The radar phase and magnitude from the key physiological sites show periodic signals corresponding to the cardiac cycle, while signals from non-physiological regions lack recognizable cardiac patterns. {\bf c,} Processing pipeline of the adaptive bin ranking algorithm. We leverage the pre-defined geometric constraints of the physiological sites, periodicity of the signal, and multi-step filtering to prioritize signal bins containing cardiac pulse features. {\bf d,} DNN architecture for multi-site PTT estimation. Our DNN architecture for PTT estimation employs a multi-branch design that jointly processes waveforms from four physiological sites (heart, neck, wrist, head) using three key components: Spatial Pooling Blocks for dimensionality reduction, Convolutional Encoder-Decoders with bi-directional LSTM for modeling the temporal relationship between cardiac waveform features, and Cross-region Fusion Modules that fuses correlated cardiac waveform features across different physiological sites.}
  \label{fig:pipeline}
\end{figure}

Simultaneously measuring and isolating cardiac signals from multiple upper-body sites using a single radar system can be challenging for two key reasons. \textit{First,} the physiological sites targeted by our system are small — the radial artery is 3~mm wide~\cite{giannattasio2005arterial,liang2023airbp}. To achieve the necessary spatial precision, we leverage a high-resolution mmWave radar system (Texas Instruments AWR2243 Cascaded Radar RF Evaluation Module~\cite{tiradar}), with 12 transmitters and 16 receivers, creating an effective array of 86 virtual antennas (\autoref{fig:overview}c). This configuration enables beamforming with an angular resolution of 1.4\degree. In contrast, prior systems~\cite{liang2023airbp,cao_hbp-fi_2024,hu_contactless_2024} leverage smaller antenna arrays with 4 to 12 virtual antennas, limiting their angular resolution. Our signal processing pipeline then spatially decomposes the received signal into range-angle bins using range FFT and beamforming (\autoref{fig:pipeline}a), applies a coarse spatial filter based on the known position of the participant and key physiological sites to classify signals as coming from the wrist, head, neck or heart. Finally, we design an adaptive bin ranking algorithm leveraging the autocorrelation properties, signal strength and periodicity, to prioritize signal bins containing the strongest cardiac pulse features.
% and ranks the signal bins by their autocorrelation properties and signal strength to identify the strongest periodic cardiac signals (\autoref{fig:pipeline}b).

\newpage
\thispagestyle{empty}

\textit{Second,} the received radar signal is a combination of reflections from multiple body regions, and environmental interference. As the pulse-induced displacements at distal arteries, such as the radial artery, are minute, approximately $\approx30~\mu m$~\cite{giannattasio2005arterial,liang2023airbp}, they can be overshadowed by noise sources within the radar’s beam diameter such as minor muscle tremors at the wrist and neck and, respiratory motions at the heart which produce larger displacements of 1--4~mm~\cite{pedersen2004breathing}. 

{To enable pulse transit time estimation in the presence of unwanted reflections, we design a DNN model to identify the periodic cardiac patterns in the radar signal and detect the timing of the key cardiac waveform features corresponding to the starting and ending timing reference for pulse transit time. The DNN model takes the phase and magnitude waveforms from the highest-ranked bins at the four key physiological sites as inputs, outputs a probability distribution of these signal features, and estimates PTT (Fig.~\ref{fig:pipeline}c). Specifically, our DNN architecture consists of three key components: a spatial pooling block, a convolutional encoder-decoder architecture, and a cross-region fusion module.  }
% To isolate the cardiac signal from unwanted reflections, we first use a bandpass filter to discard interfering signals that could correspond to muscle tremors and unwanted motion. Then to identify the key cardiac waveform features corresponding to the starting and ending timing reference for pulse transit time, 
% the phase and magnitude waveforms from the highest-ranked bins at the four key physiological sites are passed as input to our DNN model which outputs a probability distribution of these signal features, and estimates PTT (Fig.~\ref{fig:pipeline}c).
% Our DNN architecture consists of three key components: a spatial pooling block, a convolutional encoder-decoder architecture, and a cross-region fusion module. 

{Instead of relying only on the highest-ranked bins, we input a contiguous spatial region of bins centered around the four physiological sites to the DNN model. This design is motivated by the observation that the neighboring range and angle bins also contain valuable information as the pulse propagates across adjacent tissue regions. As the first DNN module to preprocess the radar signal, we design a spatial pooling block based on 1D Convolutional Neural Network (CNN) and Top-K spatial pooling filter to reduce the spatial dimensionality of the input waveforms. }

To extract features from each region's waveforms we use a convolutional encoder-decoder architecture with U-Net skip connections~\cite{UNet} and a bi-directional long short-term memory (LSTM)~\cite{yu2019review} in between them for temporal modeling. This architecture captures both local waveform characteristics and longer temporal dependencies essential for accurate detection of cardiac waveform features.

Our network uses a cross-region fusion module based on attention layers that leverages the observation that cardiac signals at different body sites are likely to share common waveform characteristics as they originate from the same cardiac cycle. By fusing information across different measurement sites, our system is able to jointly estimate PTT across multiple sites, and is more resilient to noise or signal degradation at a single site.

\subsection*{Clinical testing}

% \newpage
% \input{table_demo}

\noindent {\bf Study demographics.} We conducted a human subjects study on 35 participants in an office environment on the Carnegie Mellon University campus (Supplementary Table~\ref{tab:demo_phases}). Of these participants, 4 participants had a high body mass index (BMI), defined as BMI > 30~kg/m\textsuperscript{2}, 2 had atrial fibrillation, 2 had diabetes, and 1 had hypertension. Participants were recruited by word of mouth and an online platform for clinical study recruitment (Pitt+Me) with a mean age of 31 ± 15 years, height of 174.6 ± 8.0~cm, weight of 74.1 ± 16.3~kg, and female-to-male ratio of 0.46. All participants provided written informed consent prior to enrollment. 

%which represents the \red{first-second peak} of the carotid area IMU waveform and the foot point of the PPG waveform from the PPG sensor (\autoref{fig:pipeline}).
% The starting point of the pulse is denoted by the aortic opening (AO) in the SCG signal, or the J-peak in the BCG signal; the ending point of the pulse is measured as the systolic foot (SF) of the carotid artery pulse waveform and the foot point of the radial artery pulse waveform. 

\noindent {\bf Clinical study protocol.} Participants sat upright in a chair positioned next to a table. The radar device was placed underneath the table, positioned approximately 10~cm below the wrist, and angled 10\degree\ towards the chest. This positioning resulted in the radar being approximately 50~cm from the chest and 65~cm from the neck and 70~cm from the head. Participants were instructed to rest their right arm on the table above the radar. To obtain a ground truth reference for PTT, contact-based IMU sensors were secured using straps at the apex of the heart, and the mastoid area behind the ear, and attached using medical tape at the base of the carotid artery above the left clavicle; a PPG sensor was affixed to the radial artery on the wrist using medical tape (Supplementary Fig.~\ref{fig:ref_sensors}). The ground truth reference for blood pressure was obtained using a commercial blood pressure cuff (Omron BP7900~\cite{OmronBP7900}). Testing was conducted with participants wearing their regular clothing, which ranged from lightweight T-shirts to thicker sweaters.

% \input{table_demo}

% \red{TODO: why are we asking them to exercise, and why did we choose cycling? -- To induce variations in PTT, various non-invasive approaches can alter PTT, including cold water immersion and breathing maneuvers, where exercise produces the most substantial PTT changes~\cite{di2021multi}. As Cardiopulmonary Exercise Testing (CPET) is considered the gold standard for the comprehensive evaluation of cardiovascular responses~\cite{wibmer_pulse_2014}, we chose cycling as the method to induce the necessary hemodynamic changes for robust system validation.} 
The study consisted of two phases: a development phase, during which data was collected to create the signal processing and neural network algorithms, and an evaluation phase, in which the system was prospectively tested on a new cohort of participants. Each participant's data collection session was divided into multiple measurements, with each measurement involving simultaneous recording of PTT using both the mmWave radar system and the contact-based reference sensors for three minutes. Between measurements, the contact-based sensor was removed and then reattached to the body. To induce large variations in PTT and BP, we had participants engage in an exercise protocol (Supplementary \autoref{fig:timeline}) that involved cardiopulmonary exercise by cycling on a stationary bike as dynamic exercises that require substantial metabolic demand is a commonly employed intervention that is known to trigger significant changes in PTT and BP~\cite{mukkamala2015toward}.

During the training phase, we recruited healthy subjects ($n=22$) with a mean age of \(25.6 \pm 6.2\) years, height of \(175.4 \pm 8.2\) cm, weight of \(73.0 \pm 16.5\) kg, and female-to-male ratio of 0.29 (Supplementary Table~\ref{tab:demo_phases}). A subset of participants ($n=10$) completed a preliminary exercise protocol involving three measurements at rest, followed by five minutes of stationary cycling at 100\% intensity, and three additional post-exercise measurements (Supplementary \autoref{fig:timeline}). The remaining participants ($n=14$), including two subjects who participated in the preliminary protocol, underwent a full experimental protocol involving a wider range of exercise intensities (Supplementary \autoref{fig:timeline}). The full protocol began with two initial measurements at rest, followed by stationary cycling at sequentially adjusted exercise intensities, 50\%, 75\%, 100\%, 75\%, and 50\%. Two measurements were taken between each exercise intensity level, with two final measurements at the conclusion of the protocol, {resulting in 14 measurements in total for the full experimental protocol}. Throughout all measurements, participants remained static while breathing normally. 

In the evaluation phase of the study, participants ($n=13$) were recruited both through Pitt+Me and by word of mouth at Carnegie Mellon University. These subjects completed the full experimental protocol, with blood pressure recorded at the beginning of each measurement using the commercial blood pressure cuff. The mean age of the subjects was \(40.8 \pm 20.9\) years, with a mean height of \(173.3 \pm 7.9\) cm, weight of \(76.0 \pm 16.5\) kg, and a female-to-male ratio of 0.86 (Supplementary Table~\ref{tab:demo_phases}). {Among the 182 measurements collected in the evaluation phase (13 participants $\times$ 14 measurements), 24 of them were excluded due to transient signal quality issues with the experimental setup. We evaluate the system performance of pulse transit time on the remaining 158 measurements, and diastolic blood pressure estimation performance on 119 measurements (3 measurements of each user are used for calibration).}

\begin{figure}
    \centering
    \includegraphics[width=0.7\linewidth]{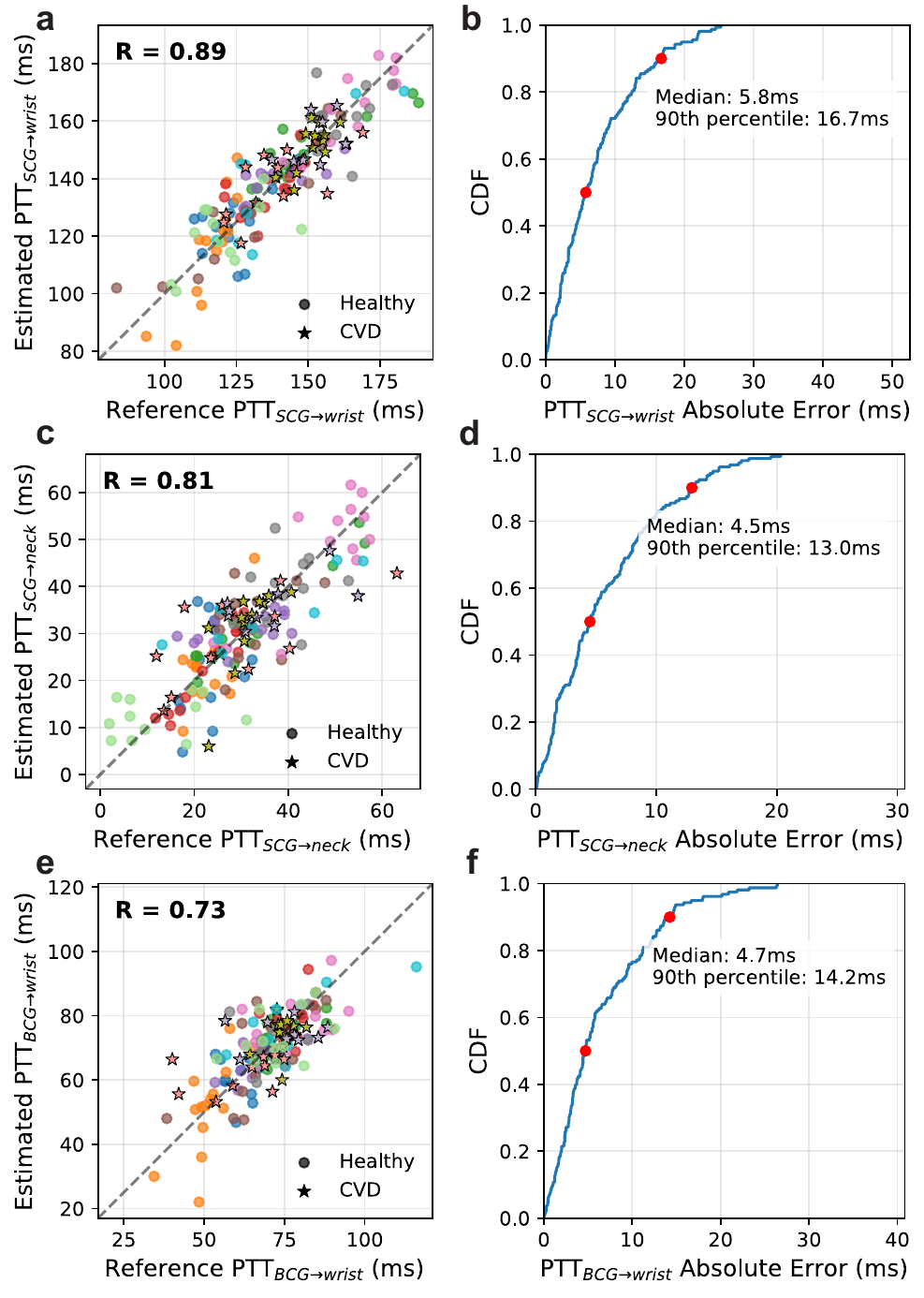}
    \caption{{\bf Comparison of PTT estimates from the mmWave radar system against contact-based reference sensors.} {\bf a, c, e} Correlation plots of PTT estimates and {\bf b, d, f} Cumulative Distribution Function~(CDF) plots of PTT errors for $\pttscgwrist$, $\pttscgneck$, and $\pttbcgwrist$. Data is shown for 13 participants (visualized in different colors) across 158 sessions in total.}
    \label{fig:main_result}
\end{figure}

\noindent {\bf Clinical testing of pulse transit time.} 
% \red{Among the PTT pathways that our system can derive from four arterial points, we report results for the three most commonly used paths: $\pttscgwrist, \pttscgneck, \pttbcgwrist$~\cite{wang2018seismo, geng2023contactless, yousefian2019potential}.} 
% Fig.~\ref{fig:main_result} presents the performance of the radar system compared with the contact-based reference sensors. Specifically, $\pttscgwrist$ exhibited a correlation coefficient of \(R=0.89\) with a median difference (MD) of 5.8 ms between reference and radar measurements, $\pttscgneck$ yielded \(R=0.81\) with an MD of 4.5 ms, and $\pttbcgwrist$ produced \(R=0.73\) with an MD of 4.7 ms. 
The PTT values computed by the mmWave radar system was compared against the contact-based reference sensors for the heart-to-wrist, heart-to-neck, and head-to-wrist physiological pathways. 
For each measurement, we use the median PTT value derived from both the mmWave radar system and reference sensors for evaluation.
We note that we do not analyze the head-to-neck ($\pttbcgneck$) pathway as the distance between the physiological points is substantially shorter than the other pathways. The correlation coefficient for $\pttscgwrist$, $\pttscgneck$, and $\pttbcgwrist$ was $R=0.89$, $R=0.81$, and $R=0.73$ respectively (Fig.~\ref{fig:main_result}a, c, e). The median absolute error for $\pttscgwrist$, $\pttscgneck$, and $\pttbcgwrist$ was 5.8, 4.5, and 4.7~ms respectively, while the 90th percentile absolute error was 16.7, 13.0, and 14.2~ms respectively (Fig.~\ref{fig:main_result}b, d, f). Related works~\cite{cho2021measurement} leveraging ultra-wideband radar to compute PTT report mean errors of 56--58~ms, while related works~\cite{buxi2016blood} computing PAT (pulse arrival time), which is the sum of PTT and the pre-ejection period~\cite{mukkamala2015toward,geddes1981pulse}, report a median error of 33~ms.

With respect to the scale of PTT total variations (the differences between the maximum and minimum PTT measured by contact-based reference sensors), the relative median PTT errors for $\pttscgwrist$, $\pttscgneck$, and $\pttbcgwrist$ were 5.5\%, 7.3\%, and 5.8\% respectively, while the 90th percentile relative errors were 15.9\%, 21.2\%, and 17.4\% respectively. We note that out of the three PTT pairs, the heart-to-neck distance is the shortest. As such $\pttscgneck$ exhibits the highest relative error, but the lowest absolute error.

% \red{TODO: repeatability/test-retest/variability analysis -- [percentage, RMSE]}
% \red{TODO: peak find rate at different sites -- discuss why the R varies at different PTT pairs}

Next, we evaluate our algorithm's ability to detect the key cardiac waveform features used to compute PTT: the aortic opening (AO) in the SCG signal, the J-peak in the BCG signal, and the waveform foots from the radial artery (wrist) and mastoid area (head). The median detection error was 10~ms for the aortic opening in the SCG signal, 14~ms for the J-peak in the BCG signal, and 12~ms and 14~ms for the waveform foot at the radial artery and the carotid artery respectively. {The higher detection error observed at the mastoid area correlates with the reduced performance of the $\pttbcgwrist$ measurements compared to $\pttscgwrist$, as shown in Fig.~\ref{fig:main_result}. This performance difference can be attributed to the anatomical characteristics of the mastoid region, which has a larger surface area than other measurement sites, and is more susceptible to motion artifacts during seated measurements due to the head's distance from the body's center of mass. }
% These involuntary motions can introduce additional noise that can obscure the BCG signal and affecting the accuracy of our cardiac waveform feature detection algorithms.

{To evaluate the repeatability of our system, we analyzed consecutive PTT measurements taken at rest prior to exercise intervention. For each subject, we compared the PTT values between two baseline measurements. The mean absolute difference (MAD) between successive measurements was 3.9 $\pm$ 1.7~ms for $\pttscgwrist$, 3.1 $\pm$ 2.3~ms for $\pttscgneck$, and 3.7 $\pm$ 2.1~ms for $\pttbcgwrist$ across all participants. In comparison, the MAD between the measurements obtained from contact-based reference sensors are 3.3 $\pm$ 2.1~ms, 3.2 $\pm$ 3.0~ms, and 4.3 $\pm$ 4.1~ms respectively. 
Prior work reports the mean and standard deviation of PTT fluctuations for subjects at rest is 14 $\pm$ 5~ms~\cite{drinnan_relation_2001}}.

\begin{figure}
    \centering
     \includegraphics[width=0.7\linewidth]{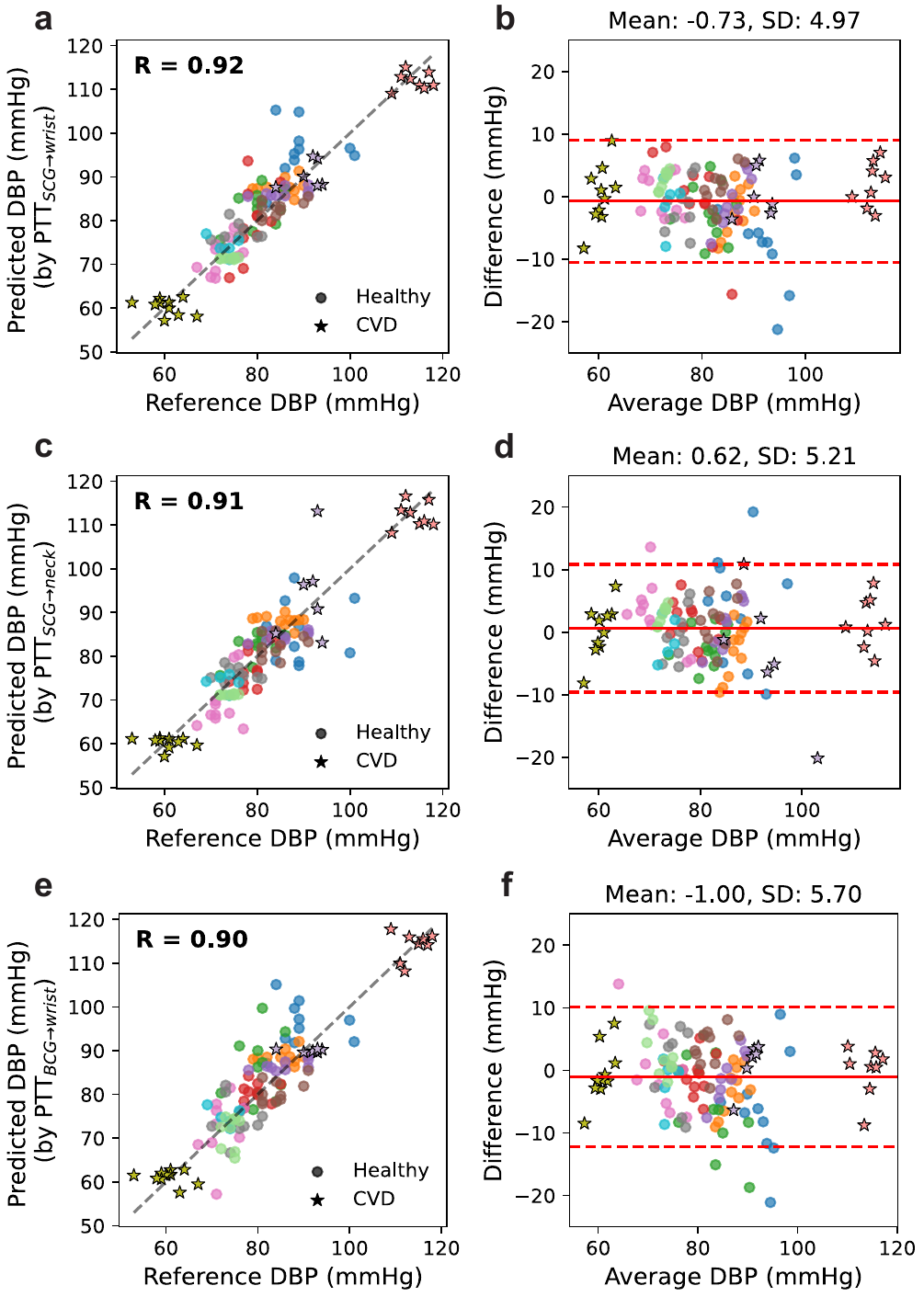}
    \caption{{\bf Comparison of diastolic blood pressure estimates from the mmWave radar system against a commercial blood pressure cuff.} {\bf a, c, e} Correlation plots and {\bf b, d, f} Bland-Altman plots of DBP for $\pttscgwrist$, $\pttscgneck$, and $\pttbcgwrist$. Data is shown for 13 participants (visualized in different colors) across 119 sessions in total. In the Bland-Altman plot, the solid line represents the mean error and the dotted lines represent the 95\% limits of agreement.}
    \label{fig:bp_result}
\end{figure}

% \red{TODO: citation for why we do DBP and not SBP --- PTT measurements using waveform foot-detection during exercise primarily capture changes in arterial stiffness and vascular tone - physiological properties more directly linked to DBP~\cite{hall_guyton_2016}}. 
% \noindent \red{TODO: Rationale for why only DBP and not SBP}\\
\noindent {\bf Clinical testing of diastolic blood pressure.} 
We also evaluated the performance of the mmWave radar system against a commercial blood pressure cuff at estimating diastolic blood pressure (DBP). To model the relationship between an individual's PTT measurements with DBP, we employed a subject-specific regression model~\cite{wibmer_pulse_2014} for calibration. For each participant, we used three recordings for calibration: one randomly selected resting-state measurement and two randomly selected post-exercise measurements. This randomization approach reduces selection bias and ensures the calibration captures a diverse range of physiological states. The correlation coefficient for $\pttscgwrist$, $\pttscgneck$, and $\pttbcgwrist$ was $R = 0.92$, $R = 0.91$, and $R = 0.90$ respectively (Fig.~\ref{fig:bp_result}a, c, e). Bland-Altman analysis~\cite{bland1986statistical} was performed for DBP estimated from $\pttscgwrist$, $\pttscgneck$, and $\pttbcgwrist$ and demonstrated a bias error of -0.73, 0.62, and -1.00~mmHg, respectively, with 3, 5, and 5 of 119 measurements falling outside the 95\% agreement limits respectively (Fig.~\ref{fig:bp_result}b, d, f). The mean absolute error for DBP estimated by our radar system from $\pttscgwrist$, $\pttscgneck$, and $\pttbcgwrist$ was 3.85, 4.03, and 4.37~mmHg, respectively, while for the contact-based reference sensors the MAE was 3.09, 3.19, and 3.62~mmHg respectively. Across these three PTT pairs, the MAE of our system is within 0.75--0.84~mmHg compared with the estimation derived from the contact-based reference sensors.

The DBP measurements derived from our radar system demonstrated performance that meets clinical guidelines for blood pressure monitoring devices. Our system achieved a mean error of $-0.73$, $0.62$ and $-1.00$~mmHg and standard deviation of errors of $4.97$, $5.21$ and $5.70$~mmHg based on estimates from $\pttscgwrist$, $\pttscgneck$, $\pttbcgwrist$ respectively. This is within the acceptable range of the FDA's Association for the Advancement of Medical Instrumentation guidelines for non-invasive sphygmomanometers~\cite{aami} which allow mean errors up to 5~mmHg and standard deviation up to 8~mmHg. Furthermore, when assessed using the British Hypertension Society (BHS) criteria~\cite{obrien_british_1990}, our system's performance met the requirements for Grade A classification for all three PTT pairs~(Supplementary Fig.~\ref{fig:bhs}). Specifically, 74.8\%, 72.3\%, and 62.2\% of measurements had an error of $\leq$ 5~mmHg for the $\pttscgwrist$, $\pttscgneck$, and $\pttbcgwrist$ PTT pairs respectively; 97.5\%, 95.0\%, and 94.1\% were within $\leq$ 10~mmHg; 97.5\%, 98.5\%, and 97.5\% were within $\leq$ 15~mmHg. This performance achieves a Grade A classification requires that at least 60\% of measurements be $\leq$ 5~mmHg, 85\% be $\leq$ 10~mmHg, and 95\% be $\leq$ 15~mmHg.

\renewcommand{\arraystretch}{1.5} % Adjust this value to make rows taller

% \noindent We performed a subgroup analysis to assess whether demographic factors affected the accuracy of our radar-based PTT measurements from the testing phase dataset. ~\autoref{fig:subgroup_analyze} illustrates the PTT error distributions across subject subgroups categorized by BMI~(a), age~(b), sex~(c), race~(d), and cardiovascular disease status~(e). The PTT ME remained below 10~ms across all demographic categories for each of the three measurement pathways ($\pttscgwrist, \pttscgneck, \pttbcgwrist$). Subjects with higher BMI ($\geq$28~kg/m\textsuperscript{2}) exhibited increased ME within $xxx-xxxms$ among three PTTs where $\pttbcgwrist$ has the smallest ME different with ME XXX compare with BMI > 25 & < 28~kg/m\textsuperscript{2} XXX and BMI \leq 25 XXX. Age-related differences in measurement error were minimal, with older subjects~($>$60~years) showing comparable performance to younger participants. No significant differences were observed between male and female subjects or across racial groups, though the limited sample size in some categories (particularly $n=1$ for Black participants) precludes \red{definitive conclusions} about these demographic factors. Subjects with known cardiovascular disease showed only \red{marginal differences} compared to healthy subjects, \red{suggesting the potential applicability of our system across diverse clinical populations.}

\begin{figure}[H]
\centering
    \includegraphics[width=\linewidth]{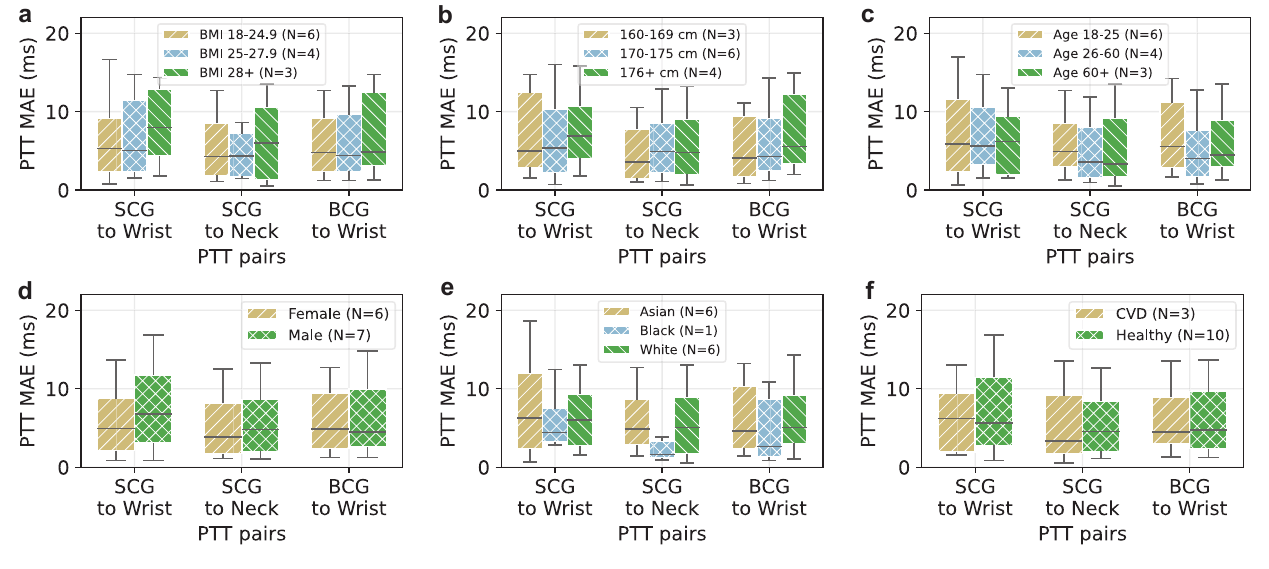}
  \vspace{-1em}
  \caption{{\bf Subgroup analysis of PTT error across study participants.} PTT Mean Absolute Error~(MAE) for $\pttscgwrist$, $\pttscgneck$, and $\pttbcgwrist$ across different {\bf a,} BMI ranges, {\bf b,} height ranges, {\bf c,} age ranges {\bf d,} female and male participants, {\bf e,} races, and {\bf f,} participants with and without cardiovascular conditions (hypertension, $n=1$ and atrial fibrillation, $n=2$). Each box in the plots shows the median error (center line), 25th and 75th percentile errors (box edges), and 10th and 90th percentile errors (whiskers).}
  \label{fig:subgroup_analyze}
\end{figure}

\noindent {\bf Subgroup analysis.} We performed a subgroup analysis to evaluate the performance of our radar-based PTT measurement system across different demographic groups (Fig.~\ref{fig:subgroup_analyze}). \\
\textit{Body mass index.} For BMI subgroups, subjects with $\mathrm{BMI} \geq 28~\mathrm{kg/m^2}$ showed median PTT errors of 8.0, 6.0, and 4.9~ms for the $\pttscgwrist$, $\pttscgneck$, and $\pttbcgwrist$, respectively. In comparison, subjects with $25 - 27.9~\mathrm{kg/m^2}$ demonstrated median errors of 5.1, 4.4, and 4.4~ms respectively, while subjects with $\mathrm{BMI} \leq 25~\mathrm{kg/m^2}$ showed median errors of 5.3, 4.3, and 4.8~ms respectively for the same PTT pairs. \\
\textit{Height.} For subjects with a height from 160-169~cm, the median error across the three cardiac pathways ranged from 3.61 to 4.96~ms. For the height group greater than 170~cm, the median error ranged from 4.80 to 6.91~ms across all pathways.\\
\textit{Age.} Comparison of PTT error by age subgroups show that subjects aged $18 - 25$ years had median PTT errors of 5.9, 4.9, and 5.6~ms across the three PTT pairs. Subjects aged $26-60$ years showed median errors of 5.6, 3.6, and 4.0~ms, while subjects greater than $ 60$ years had median errors of 6.2, 3.3, and 4.5~ms respectively. \\
\textit{Sex.} When comparing PTT measurement performance by sex, female subjects demonstrated median errors of 4.9, 3.8, and 4.8~ms for the $\pttscgwrist$, $\pttscgneck$, and $\pttbcgwrist$ PTT pairs, respectively. Male subjects showed median errors of 6.8, 4.8, and 4.5~ms respectively for the same PTT pairs respectively. \\
\textit{Race.} Analysis by race indicated that Asian subjects had median PTT errors of 6.3, 4.9, and 4.6~ms across the three PTT pairs respectively. The single Black subject in our cohort showed median errors of 4.4, 1.6, and 2.6~ms respectively, while White subjects demonstrated median errors of 6.0, 5.0, and 5.1~ms respectively. \\
\textit{Cardiovascular conditions.} Finally, we examined differences between subjects with cardiovascular disease (hypertension, $n=1$ and atrial fibrillation $n=2$) and healthy subjects. Those with cardiovascular disease had median PTT errors of 6.2, 3.3, and 4.5~ms for the $\pttscgwrist$, $\pttscgneck$, and $\pttbcgwrist$, respectively, while healthy subjects showed median errors of 5.7, 4.5, and 4.8~ms, respectively.

\begin{figure}[H]
\centering
\includegraphics[width=\linewidth]{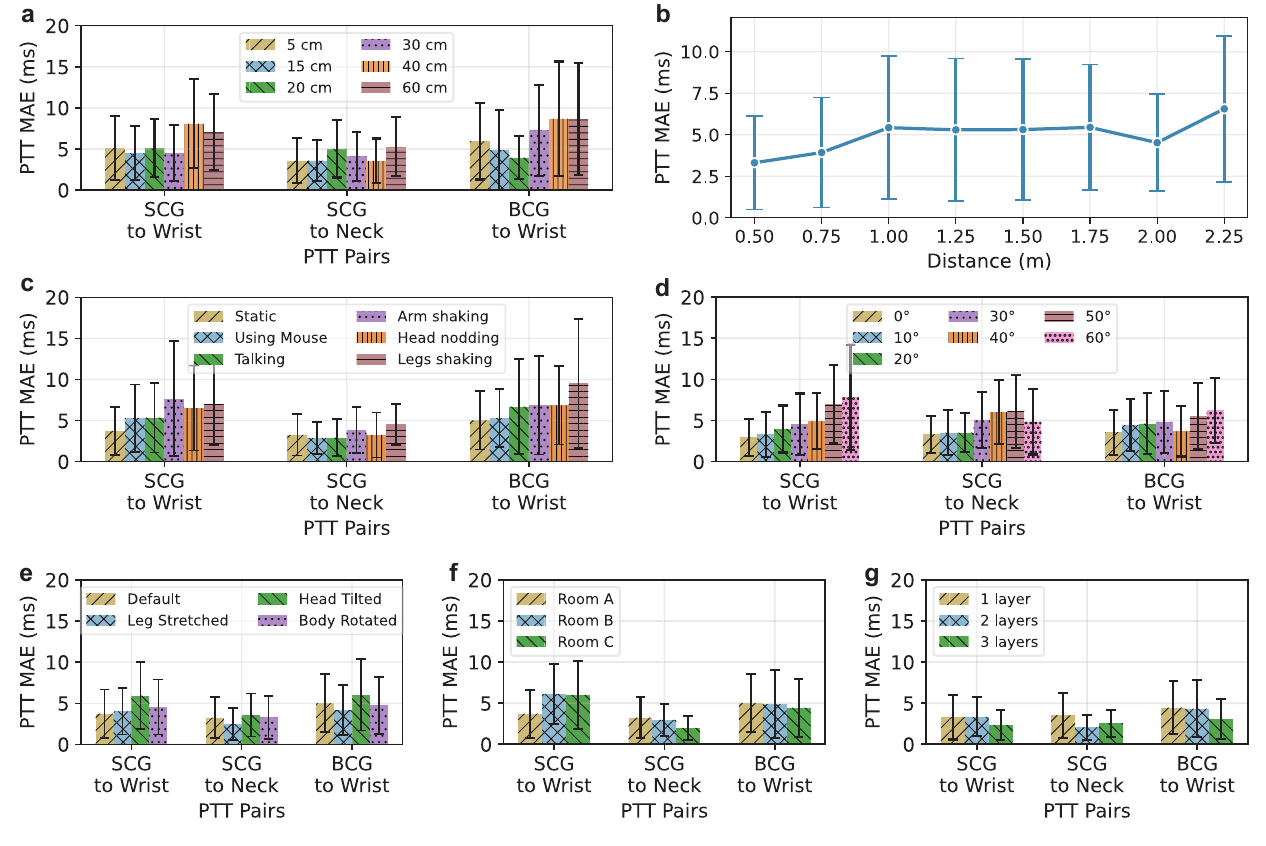}
  \vspace{-1em}
  \caption{\red{{\bf Benchmark testing across different scenarios.} The system was evaluated across different {\bf a,} radar heights, {\bf b,} radar distances, {\bf c,}  participant movements, {\bf d,} radar tilt angles, {\bf e,} participant postures, {\bf f,} testing environments, and {\bf g,} clothing layers. The bar plots show the PTT mean absolute error across the pulses in a 2--3 minute session, with the error bars showing the standard deviation of the errors. The line plot shows the $\pttscgneck$ errors.}}  \label{fig:bm}
\end{figure}

\noindent {\bf Benchmark testing.} 
% We performed benchmark experiments to evaluate the performance of our radar system across a range of experimental conditions and environmental scenarios (Fig.~\ref{fig:bm}).
{We performed benchmark experiments on a single participant to evaluate the performance of our radar system across a range of experimental conditions and environmental scenarios (Fig.~\ref{fig:bm}).} 
We systematically evaluated design and environmental factors that could affect system accuracy, including the radar’s distance to the participant, radar tilt angle, number of clothing layers, fidgeting movements by the participant, and different indoor environments. 
{For each experimental setting, we collected measurements over 2--3 minutes and calculated the per-pulse median absolute PTT errors.}
% All tests were conducted using the model trained on the combined Phase~1 and Phase~2 datasets. 

\noindent \textit{Effect of radar height.} We varied the radar height from 5 -- 60~cm below radial artery, as shown in Fig.~\ref{fig:bm}a, to examine its effect on system performance. This was achieved by adjusting the height of the tripod on which the radar was mounted, directly changing the distance between the radar and the radial artery. Our evaluation shows that for the PTT pairs where the wrist is the distal site ($\pttscgwrist$, $\pttbcgwrist$), the PTT error increased from 5.11 to 8.09~ms, and from 4.87 to 8.67~ms respectively. In contrast, for $\pttscgneck$ the error exhibited a smaller increase from 3.57 to 5.28~ms. 
% These results suggest that in our system, PTT pairs with distal sites that have a smaller arterial sensing area are more sensitive to changes in distance.

% For a height change \(\Delta h\) from our default setup (with initial distance \(d\) and incident angle \(\theta\)), we recalculated the effective angle as $\tilde{\theta} = \arctan\!\Bigl(\frac{q}{p}\Bigr)$, 
% where \(q = d\sin\theta\) and \(p = d\cos\theta\). The adjusted distance is $\tilde{d} = \frac{p}{\cos\tilde{\theta}}$.
% By updating the target region accordingly,
\noindent \red{\textit{Effect of radar distance.}  We evaluate the effect of distance between the radar and the participant on system performance in the setup shown in Supplementary Fig.~\ref{fig:benchmark}d. In this position when the participant is away from the desk, the neck is the distal arterial point which is visible to the radar, and so we focus our evaluation on the $\pttscgneck$ pathway. As shown in Fig.~\ref{fig:bm}b, the PTT errors is in the range of 3--5~ms, across the tested range of 0.5 to 2.25~m. This aligns with the operational distance of current commercial mmWave sensing devices~\cite{gruzewska2025uwb}.} 

\noindent \textit{Effect of movements.} We evaluated the effect o movements by the participant on system performance. As shown in Fig.~\ref{fig:bm}c, across all PTT pairs, measurement errors were the lowest at 3.72--5.02~ms when the participant remained still. \red{During common everyday movements such as talking or using a computer mouse, the PTT  errors remained within a comparable range to that of the static condition.} In the presence of fidgeting movements like arm shaking and head nodding, PTT errors increased to 3.82--7.64~ms. The largest PTT error of 9.52~ms was observed in the $\pttbcgwrist$ pair during leg shaking. This is likely due to motion artifacts propagating through the body, and affecting multiple cardiac measurement points. In contrast, the $\pttscgneck$ pair showed the smallest error of 3--5~ms under motion, and was the least affected by fidgeting movements. We do note that related work measuring cardiac signals using active sonar~\cite{wang2021using}, vision-based vibrocardiography~\cite{yan2018contact}, and inertial measurement units~\cite{he_wearable_2013} also require users to remain static to obtain optimal results, as motion artifacts can mask subtle cardiac signals. Extending the system to work in the presence of natural user motions such as fidgeting would require designing motion tracking and compensation algorithms capable of isolating cardiac signals from natural body movements.

\noindent \textit{Effect of radar tilt angle.} We assessed the impact of varying the radar’s tilt angle while keeping its height constant at {10~cm below the radial artery}. Here \(0^\circ\) corresponds to the radar facing upward towards the ceiling, and positive angles indicate a tilt toward the participant. The tilt angle was varied from \(0^\circ\) to \(60^\circ\) in increments of \(10^\circ\). Fig.~\ref{fig:bm}d shows that all three PTT pairs exhibit similar trends, with measurement errors becoming larger as the tilt angle increases. Errors remained below 5~ms for all PTT pairs at angles up to $30^\circ$ and increased to 6.07 - 7.82~ms at $60^\circ$ across all pairs. This is likely due to stronger signal reflections from the wrist and neck at smaller tilt angles.
% In this experiment, the radar was rotated about its central axis. 
% where the cardiac signal remains sufficiently strong.
% The anatomical constraints were updated according to $\theta_{\text{test}} = \theta_{\text{current}} - \Delta\theta$, where \(0^\circ\) corresponds to the radar facing upward, and positive angles indicate orientation toward the participant. 

\noindent \red{\textit{Effect of user postures.} We evaluated the effect of different natural sitting postures. These scenarios involve the participant tilting their head, stretching a leg, or rotating their torso to simulate common movements during sedentary activities. As shown in Fig.~\ref{fig:bm}e, measurement errors across all PTT pairs ranged from 2.44 - 6.05~ms under these postural variations. Compared to the initial static posture, these posture-induced variations had modest to minimal impact on system performance.}

\noindent \textit{Effect of experimental environments.} We assessed the system performance in different indoor environments. Fig.~\ref{fig:bm}f shows that the system consistently maintained PTT errors below 5~ms across multiple indoor locations, relative to Room A, where the main study was performed. This indicates that the system’s performance remains stable across different indoor environments.
% This indicates that our system can generalize across indoor environments.

\noindent \textit{Effect of clothing layers.} We investigated the impact of number of clothing layers on system performance by having subjects wear one, two, or three layers of indoor clothing, as illustrated in Fig.~\ref{fig:bm}g. Starting from the second clothing layer onwards, both the wrist and neck positions were fully covered by clothing. The resulting PTT errors stayed below 5~ms for all three layers, suggesting consistent performance despite additional clothing layers. This is consistent with prior work~\cite{geng2023contactless} that show mmWave signals can penetrate fabric and measure cardiac signals. We note that related systems that leverage optical or laser sensing technologies~\cite{niu2023full,beeckman_enhancing_2023} to measure cardiac signals, can only operate effectively on exposed skin, while acoustic sensing technologies~\cite{wang2021using} to measure heart rhythms are limited to penetrating only a single layer of clothing.

% \jiangyifei{\bf Remembered to reassign the subplot index}

\section*{Discussion}

This work represents the first system to measure multi-point pulse transit time and diastolic blood pressure using a single radar. {Prior work~\cite{zhang2017hybrid,nitzan2001difference,tanaka1998absence,cunha1997association} on BP estimation using PTT has largely focused on measuring a single PTT pair from the heart to distal sites including the finger, ear, or forehead. However, BP estimates obtained through each of these pairs may not be equivalent due to differences in hemodynamic controls which affect vessel characteristics and blood distribution at different sites~\cite{siaron_blood_2020}.} The ability to simultaneously measure PTT at multiple sites can enable a more comprehensive understanding of differences in the physiological pathways across the body. 

While blood pressure provides a standardized measure of cardiovascular health, prior work~\cite{di2021multi} has noted the value in measuring PTT alone. A marked decrease in an individual's PTT compared to their baseline could be indicative of cardiovascular risk. As such, PTT alone has been proposed as a risk marker for remote monitoring of individuals at risk of cardiac events.

Our work is focused on comparing the relationship between PTT and DBP as prior work suggests that arterial stiffness and vascular tone, key factors affecting PTT, correlate more strongly with DBP than SBP~\cite{wang2018seismo, payne_pulse_2006}. Our system leverages cardiac waveform foot-detection methods which have also been shown to correlate with changes in DBP~\cite{wang2018seismo}. We note that related studies using contact-based methods measure pulse arrival time (PAT), which measures the time interval between the ECG R peak and peripheral pulse points, is empirically more closely correlated with SBP~\cite{payne_pulse_2006, ganti_enabling_2021}.

The systolic and diastolic blood pressure estimation errors, obtained after calibration using a regression model best fitted for each subject across all data points, are presented in Supplementary \autoref{fig:sbp_compare}. For DBP, the radar system achieved a mean absolute error (MAE) of $2.62 \pm 0.82$ mmHg for $\pttscgwrist$ and $2.55 \pm 0.73$ mmHg for $\pttscgneck$, comparable to the contact-based reference sensors which demonstrated MAEs of $2.26 \pm 0.54$ mmHg and $2.36 \pm 0.58$ mmHg, respectively (Supplementary \autoref{fig:sbp_compare}a, e). For systolic blood pressure (SBP), both radar-based and contact-based approaches exhibited higher errors. The radar system yielded MAEs of $7.08 \pm 2.79$ mmHg for $\pttscgwrist$ and $6.84 \pm 3.00$ mmHg for $\pttscgneck$, while the contact-based sensors reported MAEs of $6.47 \pm 2.38$ mmHg and $6.32 \pm 2.76$ mmHg, respectively (Supplementary \autoref{fig:sbp_compare}b, f). The cumulative distribution function (CDF) plots (Supplementary \autoref{fig:sbp_compare}c, g) indicate that the error distributions for SBP estimation remain comparable between the radar system and the contact-based sensors across the entire error range. \red{In addition, Supplementary \autoref{fig:sbp_compare}d, h shows the correlation between the SBP estimates of the contact-based sensors and our radar system against the blood pressure cuff. The SBP correlation coefficient for $\pttscgwrist$ and $\pttscgneck$ was $R = 0.86$ and $R = 0.87$ respectively for our radar system, which is similar to the contact-based sensors with $R = 0.90$ and $R = 0.89$ respectively.}

Our system has the following limitations. 
First, our current design focuses on measuring physiological sites on the upper body. We do note however that our system can in principle measure the cardiac signal at arterial points on the lower body such as the femoral artery which exhibits a pulse-induced skin displacement of approximately 200~$\mu$m, which is similar to that of the carotid artery~\cite{badhwar_clinical_2024}. Extending the system to simultaneously measure arterial points on both the upper and lower body is a goal for future work and would require hardware modifications---such as increased transmission power and additional antennas along the elevation axis. These additions would not only enable whole-body pulse detection but could also accommodate a broader range of user postures beyond the current seated configuration.
% , thereby increasing the versatility and potential adoption of the technology in daily living environments.

% Second, our clinical study included a limited number of subjects with cardiovascular conditions, specifically, one participant with diagnosed hypertension and two with atrial fibrillation. While our system showed no significant performance degradation across these subjects, who represented a diastolic blood pressure range (53 - 117~mmHg), this limited sample prevents us from drawing definitive conclusions about the system's performance across diverse cardiovascular pathologies. However, comprehensive clinical validation requires future studies with larger and more diverse patient populations, particularly those with conditions affecting arterial stiffness and pulse wave propagation, such as advanced hypertension, arteriosclerosis, and neurodegenerative disorders like Alzheimer's disease. Such validation would be essential to establish the system's clinical utility across the spectrum of cardiovascular health and disease states.
Second, our clinical study included a limited number of subjects with cardiovascular conditions, specifically, one participant with diagnosed hypertension and two with atrial fibrillation. While subgroup analysis did not reveal any significant differences in PTT errors between healthy participants and those with CVD, future studies are required to assess the performance of our system across a larger number of participants with cardiovascular disease and related health conditions that affect PTT and DBP.

% The maintenance of measurement accuracy in these cases is encouraging, suggesting potential robustness across varying cardiovascular conditions. 

% Third, similar to related work measuring cardiac signals using active sonar~\cite{wang2021using}, vision-based vibrocardiography~\cite{yan2018contact}, and inertial measurement units~\cite{he_wearable_2013}, our system requires users to remain static for the duration of the recording, as motion artifacts can mask subtle cardiac signals. Extending the system to work in the presence of natural user motions such as fidgeting would require designing motion tracking and compensation algorithms capable of isolating cardiac signals from natural body movements. 

% Our current design assumes that the radar is positioned beneath the user and relies on predefined distance and angular thresholds to estimate the locations of various arterial points. For each measurement, the target pulse location need to be within the radar's direct line-of-sight and remain unobstructed by other body parts, as the radar signal can penetrate clothing but not human tissue.

\red{Finally, our proof-of-concept system was implemented on a mmWave evaluation board with an approximate cost of \$4,400. While this development platform was selected to enable rapid prototyping, in a commercial version of our design, the core algorithms would be integrated into a PCB design with a lower cost.}
% and miniaturized into an ASIC design with a lower cost. 
% This mirrors the development trajectory of commercial mmWave systems such as the Google Soli chip, where gesture recognition algorithms were initially prototyped on desktop-sized radar boxes and later miniaturized into an IC that was integrated onto Google Pixel smartphones and Nest Hub smart home devices~\cite{lien2016soli}.

% Radar pricing:
% Frontend board: https://www.mouser.com/ProductDetail/Texas-Instruments/MMWCAS-RF-EVM?qs=BJlw7L4Cy7%252bc0ksoLn3B4Q%3D%3D
% backend board: https://www.mouser.com/ProductDetail/Texas-Instruments/MMWCAS-DSP-EVM?qs=BJlw7L4Cy78S7jkNW2PN5A%3D%3D

In summary, we present a proof-of-concept system that demonstrates the feasibility of leveraging mmWave radar for multi-site estimates of PTT and DBP. Given the increasing prevalence of mmWave radar technology, our approach has the potential to enable a more accessible means of monitoring cardiovascular health without cuffs or contact-based sensors. Future studies are needed to validate the system across a broader patient population and across longer time scales.
\section*{Methods}

\subsection*{Study design} 
This study was approved by Carnegie Mellon University's Institutional Review Board \\(STUDY2024\_00000342). All studies complied with relevant ethical regulations. Participants were recruited by word of mouth from the Carnegie Mellon University student community and via Pitt+Me from the general public in Pittsburgh PA, USA. Written consent was obtained for human subjects participating in the study. Randomization was not applicable and investigators were not blinded.

Healthy participants above the age of 18 were eligible for the study. Exclusionary criteria include pregnant women and nursing mothers. Participants were screened via inclusion and exclusion criteria prior to the study.

% Exclusionary criteria include a history of alcohol or substance abuse, kidney failure or liver disease, unusual pain sensitivity or lack of sensitivity, chronic myofascial, inflammatory, neuropathic pain, use of medication known to interfere with naloxone, currently on opioids and pregnant women and nursing mothers. 

% inclusion, exclusion crtieria

\subsection*{Measurement setup} 
The mmWave radar contains a 2D array of 12 transmitter antennas with $T_\theta=9$ and $T_\psi=3$ antennas in the azimuth and elevation angle respectively, and 16 receiver antennas with $R_\theta=16$ antennas in the azimuth angle respectively. After eliminating duplicate virtual antenna locations, a total of 86 virtual antennas remain. The radar transmits at a power of 13~dBm (20~mW) which is within the typical range for Wi-Fi routers~\cite{power}. The radar was mounted on a tripod positioned approximately 10~cm below the wrist and angled 10$\degree$ toward the chest.

\subsection*{Measuring cardiac signals using mmWave radar} 
The mmWave radar transmits a chirp with linearly increasing frequency from $f_0=77$~GHz to $f_1=81$~GHz where the detailed radar configuration can be found in Supplementary Table~\ref{tab:radar_config}. The received signal contains reflections from the human body, which capture the subtle changes in the distance between the antenna and the body which are modulated by the cardiac signals. However, as the received signal is a combination of reflections from different parts of the body which contain multiple cardiac signals and motion, as well as changes in the environment, we next spatially decompose the reflections into different range and angle bins, and identify the bin corresponding to the physical location where a target cardiac signal is the strongest.

\subsection*{Algorithm to localize cardiac signals}

% \swarun{Can you walk through an architectural figure to explain the steps?} 
The goal of this step is to spatially isolate reflections caused by a target cardiac signal and minimize interference from other cardiac signals and motion from the body, as well as environmental sources. The localization process involves four steps:
% In the first step, we spatially separate the signal into a 2D space by using FMCW range-FFT and beamforming~\cite{liang2023airbp,lien2016soli,ha2020contactless}. In the second step, we leverage prior knowledge about the known position of participants relative to the device and the relative distances between arterial points, to coarsely localize the region of interest. In the third step, we identify periodic signals corresponding to the cardiac cycle. In the fourth step, we refine our estimate to discard any interfering periodic reflections that could correspond to muscle tremors and other unwanted signals. \swarun{Could you add some ''take-way'' boxes with the key solutions to technical challenges? Right now, this reads like a recipe-book and I am missing what are the novel/unique parts that are solutions to technical challenges. Some obvious challenges: how do you tell apart PTT from random body movement (idea: leveraging periodicity, known distance between points of interest). How do you know where to look for the signal (idea: leveraging body geometry).  }

\noindent {\bf Signal spatial isolation.} 
To extract cardiac-related reflections, we leverage FMCW radar processing to decompose signals based on their distance and use beamforming to separate signals based on their angle. The FMCW radar continuously transmits chirp signals with linearly increasing frequency. The receiving antennas capture the reflected signal from objects in the environment and mix the transmitted signal to produce the intermediate frequency (IF) signal. The transmitted signal \( x_{\text{TX}}(\tau) \) and received signal \( x_{\text{RX}}(\tau) \) at time \( \tau \) within a chirp period \( T \) can be formulated as
$x_{\text{TX}}(\tau) = A_{\text{TX}} \cdot e^{j(2\pi f_0 \tau + \frac{\pi B}{T} \tau^2)}$ and $x_{\text{RX}}(\tau) = A_{\text{RX}} \cdot e^{j(2\pi f_0 (\tau - t_d) + \frac{\pi B}{T} (\tau - t_d)^2)}$ where \( A_{\text{TX}} \) and \( A_{\text{RX}} \) denote the signal amplitude of transmitted and received signals, \( f_0 \) is the starting frequency, \( B \) is the bandwidth, and \( t_d \) is the round-trip time delay between transmitting and receiving. The IF signal \( x(\tau) \) is obtained by mixing the received signal with the conjugate of the transmitted signal, given by \( x(\tau) = x_{\text{RX}}(\tau) x_{\text{TX}}^*(\tau) \). {The analog-to-digital converter on the radar is configured to sample \( N \) points per chirp, where the discrete time is given by \( t = \frac{N\tau}{T} \). By applying the Fast Fourier Transform (FFT) in the discrete-time domain $\mathcal{F}$, we obtain the frequency components \( X_n = \mathcal{F}[x[t]] \), where \( n \) indicates the range bin index from $0$ to $N$. }

% The discrete sampling interval is given by \( t = 1/f_s \), where \( f_s \) denotes the chirp sampling frequency. The range resolution is defined as \( \frac{c}{2B} \), where \( c \) is the speed of light and \( B \) is the signal bandwidth.

In addition to separating reflections by range, we apply receiver-side beamforming to spatially isolate reflections in the angular dimension. Specifically, for a radar system with multiple receiving antennas, we perform a spatial FFT across the antenna array to separate reflections along the azimuth angle. This process is expressed as $X_{n, \theta} = \mathcal{F} [X_n[r]]$, where \( \theta \) represents the azimuth angle bin index, and \( r \) is the index of the receiving antenna.

% \red{By combining range-FFT and beamforming, we create a 2D spatial mapping of reflections. This approach enable us to focus on regions where cardiac signals are likely to originate while filtering out reflections from other sources.~\cite{liang2023airbp}}

\noindent {\bf Localizing using known anatomical constraints.} We leverage our knowledge of human anatomy and the relative positioning of the radar to constrain our search space for cardiac signals. Based on our experimental setup with the radar positioned below the subject’s body, we can spatially map the relevant anatomical regions as shown in Fig.~\ref{fig:pipeline}. 
% \red{In this configuration, the radar is directed approximately 10~cm toward the wrist at $0^\circ$, 55 cm toward the heart at $-40^\circ$, 70~cm toward the head at $-15^\circ$, and 65 cm toward the neck at $-20^\circ$.}

In the configuration the distance between the radar and the wrist, heart, neck, and head is 10, 50, 65, and 70~cm respectively. We configure the radar with $N=128$ samples per chirp, yielding a range resolution of 4.22~cm. This allows us to define specific range bins for each anatomical target: $n_{wrist} \in [4.22,16.88]$~cm for the radial artery, $n_{heart} \in [37.98,54.86]$~cm for the apex of the heart, $n_{carotid} \in [50.64,67.52]$~cm for the carotid artery, and $n_{mastoid} \in [59.08,75.96]$~cm for the mastoid area. Similarly, by applying FFT in the azimuth dimension with a size of 128, we achieve an angular resolution of $1.4^\circ$. This allows us to constrain the angular search space enough to specific sectors for each anatomical target: $\theta_{wrist} \in [-20,20]^\circ$ for the radial artery, $\theta_{heart} \in [-60,-20]^\circ$ for the apex of the heart, $\theta_{carotid} \in [-40,0]^\circ$ for the carotid artery, and $\theta_{mastoid} \in [-35,5]^\circ$ for the mastoid area.

These anatomically informed constraints reduce the search space substantially, allowing our system to focus on regions most likely to contain the cardiac signals of interest while filtering out irrelevant reflections from other body regions or the environment.

\noindent {\bf Identifying periodic signals.} The goal in this step is to identify the range and angle bin where the periodicity of a target cardiac signal is maximized. {To do this, we first compute the phase values $\phi_{n,\theta}[t]$ for each range and angle bin.} {Next, we apply a bandpass filter on the phase signals to remove breathing and body motion artifacts, using different frequency ranges for each measurement site:  0.5--100~Hz at the wrist, 0.5--10~Hz at the heart and neck, and 0.7--4~Hz at the head.} {We then detrend~\cite{detrend} the phase signal to remove linear trends and center it at zero. Furthermore, we apply the differentiator filter~\cite{diff,zhao2016emotion} to sharpen the cardiac waveform features from the heart and neck.} 
%As the phase measurements can be affected by noise in the environment, we then detrend~\cite{detrend} the signal to remove any linear trends in the data.

Next, we compute the autocorrelation function (ACF) of the phase signal in all range and angular bins to detect repeating patterns that reflect the periodic nature of the cardiac cycle: {$ACF[m] = \sum\limits_{t=0}^{C} \phi_{n,\theta}[t] \phi_{n,\theta}[t+m]$, where $C$ is the number of chirps in the recording and $m$ is the time lag.} The ACF quantifies how well the signal aligns with a time-shifted version of itself. {We extract the peaks in the ACF where the time lag is greater than zero, denoted as $[p_1,p_2,\ldots,p_i,\ldots]$ at lags $[lag_1,lag_2,\ldots,lag_i,\ldots]$. For each bin, we calculate the maximum magnitude of the peaks in the ACF with time lags correspond to the typical frequency range of the heart rhythm, i.e. $lag_i/f_s$ is in the range of 0.7--3~Hz (48--180~bpm) where $f_s$ is the number of chirps per second. We then sort the bins based on the magnitudes to identify the signal with the strongest periodicity.}
% For a cardiac signal, the ACF generates peaks $[p_1,p_2,\ldots,p_i,\ldots]$ at lags $[lag_1,lag_2,\ldots,lag_i,\ldots]$ where $i$ represents the number of cardiac cycles in the recording. 
% The first peak, $p_1$, occurs at a lag of 0, indicating perfect self-correlation, while the peak, $p_i$, corresponds to the signal aligning at $lag_i$ with a periodicity of $f_p[i] = \frac{lag_i}{fs}$ where $fs$ is the chirp sampling rate.

% To identify valid cardiac signals, we examine if $f_p[i]$ at the range of {0.7--3~Hz (48--180~bpm)}, which is the typical range for a possible heart rhythm. Range and angular bins that do not meet this criterion are discarded. The remaining bins are sorted based on the amplitude of the peak $p_i$ to identify the signal with the strongest periodicity.

\noindent {\bf Filtering interference from non-cardiac reflections.} We observe that while the top bin selected in the previous step has the strongest periodicity, it may not be due to the target cardiac signal, but may be due to interference from periodic reflections. 
%We observe this to be the case for the signal recorded at the chest and the radial artery.

{To identify the bin where the SCG, BCG and carotid artery signal originates, we select the top 40 bins and re-rank them based on signal power. Since the signal from target area exhibits prominent displacement, the bin with the highest power is identified as the most likely source of the signal.}

% we compute the second derivative of the top 40 selected phase signals to derive their corresponding SCG signals. Since SCG measures acceleration and the phase signal is proportional to displacement, taking the second derivative converts displacement to acceleration. As we do not have an analytical expression of the radar phase signal, we leverage a second-order differentiator numerical method which is robust to noise~\cite{diff,zhao2016emotion}. Here, we compute $f''_0$ which is the second derivative of a sample, $f_i$ which is the phase value at the $i^{th}$ sample, and $h$ which is the time between samples: $f''_0 = \frac{4f_0+(f_1+f_{-1})-2(f_2+f_{-2}-(f_3+f_{-3})}{16h^2}$ We then rank the bins based on the power (squared amplitude) of the SCG signals and select the bin with the highest power. 

To identify the bin where the radial pulse originates, our goal is to discard outlier bins that capture periodic reflections that do not correspond to the cardiac signal. To do this, we record $lag_i$ for the top 40 bins computed in the previous step. Using a histogram with a fixed bar size of 10 data points, we identify the most frequent lag value. We then select the bin that also shares the most common lag value. 

% \subsubsection*{Problem Formulation}
% \swarun{I am missing the ``why'' of this design. I understand you are using DNNs but why? What are the alternatives? Why did you choose this design? What challenge is it solving? }
% \jiangyifei{pulse peak is not accurate, we should consider using the term "pulse event" or "the fiducial point to assess transit time"}

\subsection*{DNN model to compute pulse transit time}

\noindent {\bf Problem Formulation.} To estimate the PTT between two different body sites, we develop a temporal detection model to detect the exact timing of the key cardiac waveform features in each desired site: the aortic opening (AO) in the SCG signal, the J-peak in the BCG signal, and the waveform foots from the radial artery (wrist) and mastoid area (head).
Once the key features are identified, the PTT between two sites can be derived by calculating the time differences. Given the beamformed signal {$X_{n,\theta}[t]\in\mathbb{C}^{N\times\Theta\times T}$,} where $N, \Theta, T$ are the number of range bins, angles bins, and chirps respectively. We seek to learn a classification model $f_i$ for site $i$: $f_i\left(X_{n,\theta}[1], X_{n,\theta}[2],\dots,X_{n,\theta}[T]\right) = (y[1],y[2],\dots,y[T])$, where $y[t]\in \{0,1\}$ indicates if the timestamp $t$ has a key cardiac waveform feature, labeled using contact-based reference sensors. Instead of designing a DNN directly for binary classification, we apply label smoothing~\cite{label-smoothing} to estimate a continuous probability distribution for the key features. Specifically, we replace the binary labels at the locations of the key features with a Gaussian kernel centered at the feature's location with a maximum value of 1 and a standard deviation of 10. This formulation transforms our approach from binary classification to a temporal continuous regression model. This Gaussian representation provides more informative gradients during training, resulting in more robust model convergence than sparse binary labels.

With the range bin $\hat{n}_i$ and angle bin $\hat{\theta}_i$ with the strongest cardiac signal from site $i$ identified in previous steps, we use a contiguous spatial region of bins centered around $(\hat{n}_i, \hat{\theta}_i)$ as the inputs to the DNN. Specifically, we use 21 angle bins for heart, neck, wrist, and 5 range bins $\times$ 21 angle bins for head. We flatten the range and angles bins for head region into a single dimension. For each bin, we calculate the magnitudes and unwrapped phases (i.e. $|X_{n,\theta}[t]|$ and $\angle X_{n,\theta}[t]$) of the complex numbers, concatenate them in the spatial dimension and use as the inputs to the DNN. Given a total number of $B_i$ bins around site $i$, the model $f_i$ can be denoted as $f_i: \mathbb{R}^{2B_i\times T} \rightarrow \mathbb{R}^T$.

\noindent {\bf Loss function.} We employ binary cross entropy as our loss function to train the temporal detection model. For each site $i$, the loss $\mathcal{L}_i$ is computed between the predicted probability sequence $\hat{y}_i[t]$ and the Gaussian-smoothed ground truth $y_i[t]$:
\begin{equation}
\mathcal{L}_i = -\frac{1}{T}\sum_{t=1}^{T} \left[ y_i[t] \log(\hat{y}_i[t]) + (1-y_i[t])\log(1-\hat{y}_i[t]) \right]
\end{equation}
For our joint model that simultaneously detects the key cardiac features across all four body sites, the total loss is the sum of individual site losses (i.e. $\mathcal{L}_{\text{total}} = \sum_{i}\mathcal{L}_i$, where $i\in {\text\{heart, neck, wrist, head}\}$)

\noindent {\bf Spatial Pooling Block.} We consider a region of neighboring angle and ranges bins as the inputs of DNN since they contain valuable information as the arterial pulse propagates across the adjacent tissue region. To reduce the spatial dimensionality and suppress the noise in the raw phase and magnitude waveforms, we design a spatial pooling block as the preprocessing module of the DNN. For each body site $i$, we input waveforms from $2B_i$ spatial bins (phase and magnitude) of length $T$ samples, forming input tensor $X \in \mathbb{R}^{2B_i \times T}$. This input is processed through two consecutive 2D convolutional layers with kernel size $(1, 7)$ and padding $(0, 3)$ to preserve temporal resolution. Each layer produces 16 feature maps followed by batch normalization and ReLU activation, yielding $X' \in \mathbb{R}^{16 \times 2B_i \times T}$. We then apply Top-$K$ spatial pooling, which selects the $K$ features with the highest values across the spatial dimension, reducing dimensions to $X'' \in \mathbb{R}^{16 \times K \times T}$ where $K < B_i$. The $K$ values are set to $K_{heart}=8$, $K_{neck}=8$, $K_{head}=16$, and $K_{wrist}=4$ for the respective sites. Finally, a $1\times1$ convolution layer projects the channel dimension from 16 down to 1, producing output $\tilde{X} \in \mathbb{R}^{K \times T}$ where the spatial dimension has been effectively reduced from $2B_i$ to $K$ bins. This design distills the most relevant spatial information while maintaining sensitivity to temporal patterns in the cardiac waveform.

\noindent {\bf Convolutional Encoder-Decoder.} After spatial pooling, we implement a U-Net style convolutional encoder-decoder architecture to extract temporal features and generate sample-level probabilities of key features. The encoder consists of three consecutive blocks, each containing two 1D convolutional layers (kernel size = 7) with batch normalization and ReLU activation, followed by max pooling with stride 2. The channel dimensions progressively increase from $K$ to [32, 64, 128] to capture hierarchical features while downsampling the temporal dimension. At the bottleneck, we employ a bidirectional LSTM with 2 layers, 64 hidden units, and dropout rate of 0.2 to model long-range temporal dependencies in the signal. The decoder mirrors the encoder with three blocks that use transposed convolutions with a stride 2 for upsampling, concatenate with corresponding encoder features via skip connections, and process through two 1D convolutional layers with batch normalization and ReLU. These skip connections preserve fine-grained temporal details that might otherwise be lost during downsampling. The final output passes through two additional convolutional layers with a sigmoid activation function, producing a probability distribution $\hat{y}_i[t] \in [0,1]^T$ for each timestamp. This architecture remains consistent across all four body sites, differing only in the input spatial dimension $K$ from the spatial pooling block.

\noindent {\bf Cross-region Fusion Module.} Since pulse waves at different anatomical sites originate from the same cardiac cycle, they can exhibit temporal correlations with similar waveform features~\cite{alastruey_arterial_2023}. However, signal quality can vary across sites due to local tissue properties and measurement conditions. To exploit these relationships, we implement a cross-region fusion module between the encoder and decoder stages, after the bottleneck LSTM processes each site's feature independently. Operating on feature representations with hidden size 128, the module employs a dual attention mechanism with 4-head multi-head attention. First, cross-site attention allows each site's features to incorporate relevant information from other sites, followed by residual connections and layer normalization. Then, temporal attention is applied to each site independently, enabling focus on temporally significant patterns. This architecture enables the network to jointly model spatial relationships between anatomical sites and temporal dynamics within each site, making the system more resilient to local noise and improving PTT estimation accuracy by leveraging complementary information across all measurement sites.

\noindent {\bf Dataset Preprocessing.} We randomly divide the evaluation phase participants ($n=13$) into two equal sets. To evaluate the performance on set 1 participants, we train the model using all development phase participants plus set 2 evaluation participants, and vice versa for evaluating on set 2. Each radar measurement session lasts 80-110 seconds. We define each data sample as a 5-second segment of radar data, using a sliding window approach with a 1-second stride. This process generates approximately 10.8~k samples from development phase participants, 3.0~k samples from evaluation set 1, and 3.1~k samples from evaluation set 2.

\noindent {\bf Data Augmentation.} To improve model robustness and generalization, we implement a spatial bin dropout augmentation strategy. During training, instead of using all available spatial bins $B_i$ for each site $i$, we randomly select a subset of $B'_i$ bins in every new epoch. Specifically, we use $B'_{head}=75$ (out of 105 bins), $B'_{neck}=35$ (out of 42 bins), and $B'_{heart}=30$ (out of 42 bins), with no augmentation for wrist.
This technique forces the model to learn from different spatial combinations, preventing it from over-relying on specific bins that might have strong signals in the training data but could be less reliable in real-world scenarios. Additionally, this form of augmentation acts as a regularization mechanism to reduce overfitting and improve the model's ability to handle variable signal quality. During validation and testing phases, we use all available bins to maximize performance.

\noindent {\bf Training Procedure.} We employ a two-stage training strategy with the AdamW optimizer. In the first stage, we train the network for each individual site without cross-region fusion, using 20\% of the training data for validation. The site-specific hyperparameters are: batch size of 128 and learning rate of 0.001 for all sites, with weight decay of 0.01 for head and wrist networks and 0.0001 for heart and neck networks. We train each site-specific network for 50 epochs. In the second stage, we freeze all site-specific components and train only the cross-region fusion module with a batch size of 36, learning rate of 0.0004, weight decay of 0.0001, and for 30 epochs. This two-stage approach allows us to first build strong site-specific feature extractors and then optimize their integration.

\noindent {\bf Deriving PTT from DNN output.} To extract pulse transit times from the probability distributions of the key features produced by our model, we employ a multi-step post-processing approach. First, we identify significant peaks in the probability sequences using a peak detection algorithm with a minimum distance constraint of 300~ms between consecutive peaks and site-specific height thresholds (0.55 for heart, 0.5 for neck, 0.25 for wrist, and 0.35 for head). Once peaks are detected across all sites, we calculate PTT values between pairs of sites by taking the time difference between corresponding peaks, constraining the values to physiologically plausible ranges. To ensure measurement stability, we apply outlier removal by excluding PTT values with z-scores exceeding 2.0, followed by temporal smoothing using a moving average filter with a kernel size of 5. This processing pipeline effectively translates the model's probabilistic outputs into PTT measurements over time.

% \red{($\pttscgwrist$: 70--190~ms, $\pttscgneck$: 0--80~ms, $\pttbcgwrist$: 15--120~ms).}

\noindent {\bf PTT Performance Evaluation.} Our radar system adjusts its focus to enhance the detection of stronger cardiac signals, whereas the contact-based reference sensors remain fixed at predetermined anatomical locations throughout the measurement process. This difference in positioning introduces a spatial discrepancy between the radar’s target site and the exact placement of the reference sensors. For instance, when measuring at the neck, the contact-based sensor is positioned over the carotid artery just above the clavicle, while the radar targets a segment of the carotid artery located approximately 10 cm higher at the upper neck. Given that the pulse wave velocity typically ranges from 5 to 10~m/s~\cite{diaz_reference_2014}, and assuming a constant propagation speed between the reference sensor and the radar’s measurement site, this 10~cm displacement results in a time shift of approximately 10–20 ms. To assess the performance of radar-based PTT, we prioritize the analysis of dynamic variations rather than absolute values. Therefore, we normalize the PTT values obtained from the contact-based sensor by aligning their mean with the mean of the radar-derived PTT measurements, effectively eliminating this constant temporal offset. This normalization ensures a fair comparison of PTT fluctuations between the two modalities, as illustrated in Fig.~\ref{fig:main_result}.

\noindent {\bf Estimating Diastolic Blood Pressure from PTT.}
Several methods have been proposed for calibrating PTT with blood pressure~\cite{murakami2015non}. We select the linear regression model \( \text{DBP} = a \cdot \text{PTT} + b \)~\cite{wibmer_pulse_2014}, which is computationally efficient, requires fewer data points, and is widely used in PTT-BP calibration.

% However, due to the PTT pairs for instance, \(\pttscgneck\), are close to zero, models such as the logarithmic equation \( \text{BP} = a \ln(\text{PTT}) + b \) and the inverse equation \( \text{BP} = \frac{a}{\text{PTT}} + b \) become unstable and unsuitable for use. To address this issue,

In our blood pressure evaluation (Fig.~\ref{fig:bp_result}, Supplementary Fig.~\ref{fig:sbp_compare}), we calibrate the regression model using three data points per subject. To ensure sufficient variability in the data, one point is randomly selected when the subject is at rest, while the other two are randomly chosen from post-exercise measurements. This approach enhances the robustness of the calibration by incorporating a range of physiological states.

\subsection*{Ground truth reference measurement setup} 
{\bf Contact-based reference sensors.} To obtain a reference measurement of pulse transit time, we use the accelerometer of an inertial measurement unit (Seeed Studio XIAO nRF52840 Sense~\cite{imu}), and a PPG sensor (Electronics LLc, PulseSensor~\cite{ppg}) to measure the cardiac signals at different points along the body. The IMU is configured to have a sampling rate of 416~Hz with a linear acceleration sensitivity of 0.061~mg/LSB~\cite{accel_chip}, and the PPG sensor is configured to have a sampling rate of 500~Hz. 
We resample the IMU data to 500~Hz to make it consistent with the sampling rate of PPG and the rate of chirps of radar.
% The data from the PPG signal is interpolated to have a sampling rate of 500~Hz.

\noindent {\bf Sensor synchronization.} In our measurement setup (Supplementary Fig.~\ref{fig:ref_sensors}), an IMU and PPG sensor is packaged in a plastic case and attached to the radial artery of the subject’s right wrist and an IMU is attached to the carotid artery of neck above right clavicle using transparent wound dressing (Dimora Transparent Film~\cite{dressing}), while other two IMUs are attached to the apex of heart and mastoid area of the head using an elastic band~\cite{elastic} and velcro strap design~\cite{velcro}. All reference sensors are connected via a USB cable to a laptop for recording. The IMU and PPG at the wrist are connected to the same clock and synchronized. As the four IMU reference sensors ($IMU_{wrist}$, $IMU_{head}$, $IMU_{heart}$, $IMU_{neck}$) each have a different clock, we perform a calibration step at the beginning of each measurement to synchronize the sensor data across them and the mmWave radar. This is done by performing a common shaking motion event that is used to synchronize the sensors using the following steps:\\
\textit{Step 1.} We attach the $IMU_{neck}$ to the participant’s neck.\\
\textit{Step 2.} We attach the $PPG_{wrist}$ and $IMU_{wrist}$ to the participant’s wrist.\\
\textit{Step 3.} $IMU_{head}$ and $IMU_{heart}$ are shaken up and down along the y-axis for two seconds.\\
\textit{Step 4.} $IMU_{head}$ is attached to the head band.\\
\textit{Step 5.} $IMU_{heart}$ is held with $IMU_{neck}$ and shaken up and down along the y-axis for two seconds.\\
\textit{Step 6.} $IMU_{heart}$ is held with $IMU_{wrist}$ and shaken up and down along the y-axis for two seconds.\\
\textit{Step 7.} $IMU_{heart}$ is attached to the chest band.\\
\textit{Step 8.} $IMU_{wrist}$ is placed approximately 10~cm from the mmWave radar and shaken up and down towards the device and along the y-axis for two seconds.

To synchronize the signals, we identify the time window across a signal pair that contains the shaking motion, and use that window to compute the time lag between the recordings. To do this, we leverage a sliding window approach that segments the signal pairs into windows of duration $[wlen,wlen+step]$ where $wlen=2~s$ and $step=0.02~s$. For all window pairs, we perform cross-correlation between the two windows and find the amplitude of the maximum peak. Intuitively, the window containing shaking would contain the highest cross correlation peak because due to the synchronized shaking motion. Outside of the period of shaking, the sensors move independently and show minimal correlation. For the time window when shaking occurs, the time lag of the maximum peak is used to synchronize the signals.

We next synchronize $IMU_{wrist}$ with the mmWave radio data. We first constrain the shaking motion of wrist is 30 cm above radar. Then we select radar with the angle bin from -36 \degree to 20 \degree with the corresponding range bin. We then apply differentiator filter~\cite{diff,zhao2016emotion} on radar phase signal to align the radar signal characteristic with IMU signal. Next, we perform the same synchronization algorithm among IMU sensors with each pair of $IMU_{wrist}$ and radar range-angle bins. Finally, We select the time lag of the maximum peak with the largest peak value among selected radar bins. 

 % within range bin $n$ at distance 30~cm above the radar and the angular bin range $\theta$ at the range of $[-36,20]^\circ$.

\subsection*{Computing pulse transit time from reference sensors}
We use the PPG sensor at the wrist as a reference measure of the pulse at the radial artery, the z-axis from $IMU_{neck}$ as a reference measure of the pulse at the carotid artery, the x-axis from $IMU_{head}$ as a reference measure of BCG, and the z-axis from $IMU_{heart}$ as a reference measure of SCG.

Using our reference sensors, the $i^{th}$ PTT measurement is computed as the interval between the start of the pulse $t_{i}^{start}$ and its arrival at a peripheral arterial site $t_{i}^{end}$. In our system, we can compute the proximal point of PTT on the timing of the aortic valve opening (AO) from the SCG signal at time, or the J peak in the BCG signal. We compute the distal point of PTT at a peripheral site using the foot of the PPG signal at the radial artery at the wrist and the primary peak of the IMU signal (the foot of pulse distension) at the carotid artery at the neck~\cite{beutel_pulse_2021}. Of these two signal types (IMU, PPG), we find that the pulse waveform captured via the PPG sensor exhibits the highest SNR. As such, we first search for $t_{i}^{end}$ in the PPG signal at the wrist, then search for the corresponding $t_{i}^{start}$ and $t_{i}^{end}$ in the SCG, BCG, and neck pulse signal relative to it.

\noindent {\bf Identifying $t_{i}^{end}[Wrist]$ at Wrist from PPG Signal.} The $t_{i}^{end}$ of each pulse in the wrist PPG signal occurs at the foot of each cardiac cycle and is minimally affected by wave reflection~\cite{mukkamala2015toward}. While previous works have proposed various definitions for the foot location in a single cadiac cycle~\cite{gesche2012continuous,mukkamala2015toward,wong2009evaluation}, we use the intersecting tangent method to identify the foot point. To perform this peak selection across multiple recorded cardiac cycles, we first apply a low-pass filter with a cutoff frequency of 10 Hz to remove noise artifacts. We then compute the first derivative of the waveform to identify the point of maximum gradient. Finally, we determine the intersection of the horizontal line at the local minimum with the tangent at the maximum gradient point to find the desired foot location.

\noindent {\bf Identifying $t_{i}^{start}$ and $t_{i}^{end}$ from IMU Signals.} Next we search for the corresponding $t_{i}^{start}$ for SCG BCG signals and $t_{i}^{end}$ for carotid artery signal at neck. In the SCG signal, this is the aortic valve opening time, while in the BCG signal we use the J peak, which has been shown to have a stronger correlation with blood pressure when calculating PTT from BCG to wrist PPG~\cite{yousefian2019potential}. In the carotid artery signal, we find the first major peak following the Start of Isovolumetric Contraction (SIC). In the absence of noise, the AO peak in the SCG waveform, the J peak in the BCG signal and the major peak after SIC at the carotid artery signal are typically the highest amplitude features in the signal. However, challenges arise due to the sensitivity of the IMU signal to motion artifacts, variations in sensor placement, and differences in BMI among subjects. Consequently, the highest measured peak within a cardiac cycle may not always correspond to the target peaks. To tackle this challenge, we first perform a broad search with the reference PPG foot time $t_{i}^{end}[Wrist]$ for candidate time points to avoid selecting noise outside the $PTT$ range. Then we perform a refined search to avoid selecting neighboring peaks that obscure the desired target peaks. 
% When necessary, manual adjustments are introduced in the end to ensure accurate labeling of the target peaks.

\noindent \textit{Broad search for peaks in IMU signals.} In this step, we search for candidate time points  within periodically increasing search windows of sizes ranging from $w_{min}=100ms$ to $w_{max}=300ms$. We set the initial search window of $[t_{i}^{end}[Wrist]-w, t_{i}^{end}[Wrist]]$, where $w=w_{min}$, and progressively move the starting point backwards in time until $w=w_{max}$. The intuition for this approach is to exclude peaks resulting from unintended movements outside the expected PTT range.

We mark the timestamp of the maximum amplitude peak within the time windows as the candidate peak $\tilde{t_{i,w}}$ within the SCG, BCG, or carotid artery signal. We then sum the amplitudes of all candidate peaks in the time windows a measure of the amount of vibration energy, which we illustrate for the SCG signal: $S[w] = \sum\limits_{w=w_{min}}^{w_{max}} \sum\limits_{i=0}^{P} SCG[\tilde{t_{i,w}^{}}]$. We then find the window size which maximizes the sum of the amplitudes of the candidate starting peaks in the signal: $w^*=\argmaxA_w(S)$. We then set the coarse-grained starting time to be $\tilde{t_{i}^{}} = \tilde{t_{i,w*}^{}}$.

\noindent \textit{Refined search for $t_{i}^{}$.} Next we refine our estimate of the candidate starting time $\tilde{t_{i}^{}}$. The intuition behind this approach is that the majority of target peaks are found correctly with the window size $w^*$. To minimize the selection of neighboring peaks that may correspond to noise, we identify the most likely PTT value while applying a small threshold to ensure robustness to local PTT variance. To do this, we mark the timestamp with the maximum peak amplitude within the refined search window $[(t_{i}^{} - w^*) - \delta, (t_{i}^{} - w^*) + \delta ]$ as the final selected $t_{i}^{}$ value where the threshold $\delta=30$ms.

% \noindent \textit{Introducing Manual Adjustment for $t_{i}$.} Empirically, automated processing may fail to detect the correct target peak. We define a failure case as occurring when, for a given participant, the detected PTT within a duration $D_{fail}$ differs from other durations $D_i$ within the same subject by more than $100$ ms, or when the detected PTT produces a negative value, which is physiologically implausible. In such cases, manual adjustments are introduced to ensure accuracy. We employ a semi-automatic approach that refines time point labeling by applying a shift to $w^*$ and/or adjusting the threshold $\delta$.

% \noindent {\bf Statistical analysis.} \red{Algorithms to train, validate, and test the machine learning algorithms were performed using xxx. xxx was performed using numpy. Figures were created using matplotlib.}

\subsection*{Data availability} 
{All data necessary for interpreting the manuscript have been included. The datasets used in the current study are not publicly available but may be available from the corresponding authors on reasonable request and with permission of Carnegie Mellon University.}

\subsection*{Code availability} 
Code will be made available to public before publication.
% Code is available at \url{https://github.com/xxx}.

\section*{Acknowledgments} 
We acknowledge support from the NSF (2106921, 2030154, 2007786, 1942902, 2111751), ONR, AFRETEC, CyLab-Enterprise, and the Center for Machine Learning and Health. Any opinions, findings, and conclusions or recommendations expressed in this material are those of the author(s) and do not necessarily reflect the views of the above.

We thank our participants at Carnegie Mellon University for their willingness to participate in our study. We thank Hongyang Li, Jason Li, Ran Bi, Veronica Muriga, Yifeng Wang, and Yiwen Song for their critical and important feedback on the manuscript. We thank Chengyi Shen and Rongbing Shen for designing the 3D printed enclosure for the mmWave radio. We thank Amit Thomas for assisting with the automated labeling of data from the contact-based reference sensors. We thank Tianyi Ren for assistance with the design of the contact-based reference sensors. We thank Jihang Jiang for the support on the subject recruitment process.

\section*{Author contributions}
J.Z., K.Y., A.P., J.C. and S.K. conceptualized the study; J.Z. and K.Y. designed and performed the human subjects study, conducted the experiments, and performed the analysis under technical supervision by J.C. and S.K; J.Z., K.Y., A.P., and Y.L. developed the algorithms; G.W. designed the setup for the contact-based reference sensors; J.Z., K.Y. and J.C. wrote the manuscript; K.M. and S.K. edited the manuscript.

\section*{Competing interests}
The authors declare the following competing interests: J.C. is a co-founder of Wavely Diagnostics, Inc. The remaining authors declare no competing interests.

\section*{Supplementary Materials}
Supplementary Figure 1: Placement of contact-based reference sensors at different anatomical sites.\\
Supplementary Figure 2: Experiment protocol timeline.\\
Supplementary Figure 3: Experimental conditions and setup for benchmark testing.\\
Supplementary Figure 4: Cumulative distribution function of diastolic blood pressure error.\\
Supplementary Figure 5: Systolic Blood Pressure (SBP) and Diastolic Blood Pressure (DBP) estimation errors of contact-based reference sensor and our radar systems based on best linear fit with $\pttscgwrist$ and $\pttscgneck$.\\
Supplementary Table 1: Radar configuration parameters.\\
Supplementary Table 2: Demographic summary of participants in the evaluation cohort phase.\\
Supplementary Algorithm 1: Cardiac signal conditioning.\\
Supplementary Algorithm 2: Identification of candidate cardiac signal regions.\\
Supplementary Algorithm 3: Identification optimal cardiac signal region.

% \bibliography{refs}
% \bibliographystyle{naturemag}

% \clearpage
% \pagebreak

\pagenumbering{gobble}

\clearpage
\pagebreak

\renewcommand{\arraystretch}{0.8}
\begin{table}[h]
    \centering
    \small
    \caption{Demographic summary of participants in clinical study.}
    \label{tab:demo_phases}
    \begin{threeparttable}
    \begin{tabular}{p{3cm} p{5cm} c}
        \toprule
        % \textbf{Phase} & \textbf{Metric} & \textbf{Value} \\
        % \midrule
        %---------------- TRAINING PHASE ----------------
        \multirow{14}{*}{\shortstack[c]{\textbf{Development phase}\\($n=22$)}}
         % & \multicolumn{2}{l}{\textbf{General Information}} \\[0.5ex]
         & Age (years)       & 25.6 $\pm$ 6.2 \\
         & Height (cm)       & 175.4 $\pm$ 8.2 \\
         & Weight (kg)       & 73.0 $\pm$ 16.5 \\
         & BMI (kg/$m^2$)       & 23.6 $\pm$ 4.1 \\
         \cmidrule(lr){2-3}
         & \multicolumn{2}{l}{\textbf{Sex}} \\[0.5ex]
         & Male, $n$ (\%)    & 17 (76) \\
         & Female, $n$ (\%)  & 5 (24) \\
         \cmidrule(lr){2-3}
         & \multicolumn{2}{l}{\textbf{Race}} \\[0.5ex]
         & Asian, $n$ (\%)   & 22 (100) \\
         & White, $n$ (\%)   & 0 (0) \\
         & Black, $n$ (\%)   & 0 (0) \\
         % \cmidrule(lr){2-3}
         % & \multicolumn{2}{l}{\textbf{Cardiovascular Condition}} \\[0.5ex]
         % & Hypertension, $n$ (\%)   & 0 (0) \\
         % & AF (Atrial Fibrillation), $n$ (\%)  & 0 (0) \\
         % & Diabetes, $n$ (\%)         & 1 (5) \\
        \midrule
        %---------------- TESTING PHASE ----------------
        \multirow{14}{*}{\shortstack[c]{\textbf{Evaluation phase}\\($n=13$)}}
         % & \multicolumn{2}{l}{\textbf{General Information}} \\[0.5ex]
         & Age (years)       & 40.8 $\pm$ 20.9 \\
         & Height (cm)       & 173.3 $\pm$ 7.9 \\
         & Weight (kg)       & 76.0 $\pm$ 16.5 \\
         & BMI (kg/$m^2$)       & 25.1 $\pm$ 4.1 \\
         & SBP at rest (mmHg)       & 120 $\pm$ 24 \\
         & DBP at rest (mmHg)       & 80 $\pm$ 13 \\
         & Heart rate at rest (mmHg)       & 71 $\pm$ 9 \\
         \cmidrule(lr){2-3}
         & \multicolumn{2}{l}{\textbf{Sex}} \\[0.5ex]
         & Male, $n$ (\%)    & 7 (54) \\
         & Female, $n$ (\%)  & 6 (46) \\
         \cmidrule(lr){2-3}
         & \multicolumn{2}{l}{\textbf{Race}} \\[0.5ex]
         & Asian, $n$ (\%)   & 6 (46) \\
         & White, $n$ (\%)   & 6 (46) \\
         & Black, $n$ (\%)   & 1 (8) \\
         \cmidrule(lr){2-3}
         & \multicolumn{2}{l}{\textbf{Cardiovascular condition}} \\[0.5ex]
         & Hypertension, $n$ (\%)   & 1 (8) \\
         & AF, $n$ (\%)  & 2 (15) \\
         & None, $n$ (\%)  & 10 (77) \\
         % & Diabetes, $n$ (\%)         & 1 (8) \\
        \bottomrule
    \end{tabular}
    \begin{tablenotes}
      \footnotesize
      % \item Data are means ± SD or percentages.
      \item BMI = body mass index, SBP = systolic blood pressure, DBP = diastolic blood pressure, AF = atrial fibrillation
    \end{tablenotes}
  \end{threeparttable}
\end{table}

% \clearpage
% \pagebreak

\renewcommand{\figurename}{Supplementary Figure}
\setcounter{figure}{0}
\setcounter{table}{0}
\renewcommand{\tablename}{Supplementary Table}

% \clearpage
% \newpage

% \begin{figure}[H]
% \centering
% {\includegraphics[width=.95\textwidth]{figs/xxx}}
% \caption{\red{TODO: picture of setup with the elbow rest like sfig6 here: https://static-content.springer.com/esm/art\%3A10.1038\%2Fs42003-021-01824-9/MediaObjects/42003_2021_1824_MOESM2_ESM.pdf}}
% \label{fig:radar_setup}
% \end{figure}

\clearpage
\newpage

\begin{figure}[H]
\centering
{\includegraphics[width=.95\textwidth]{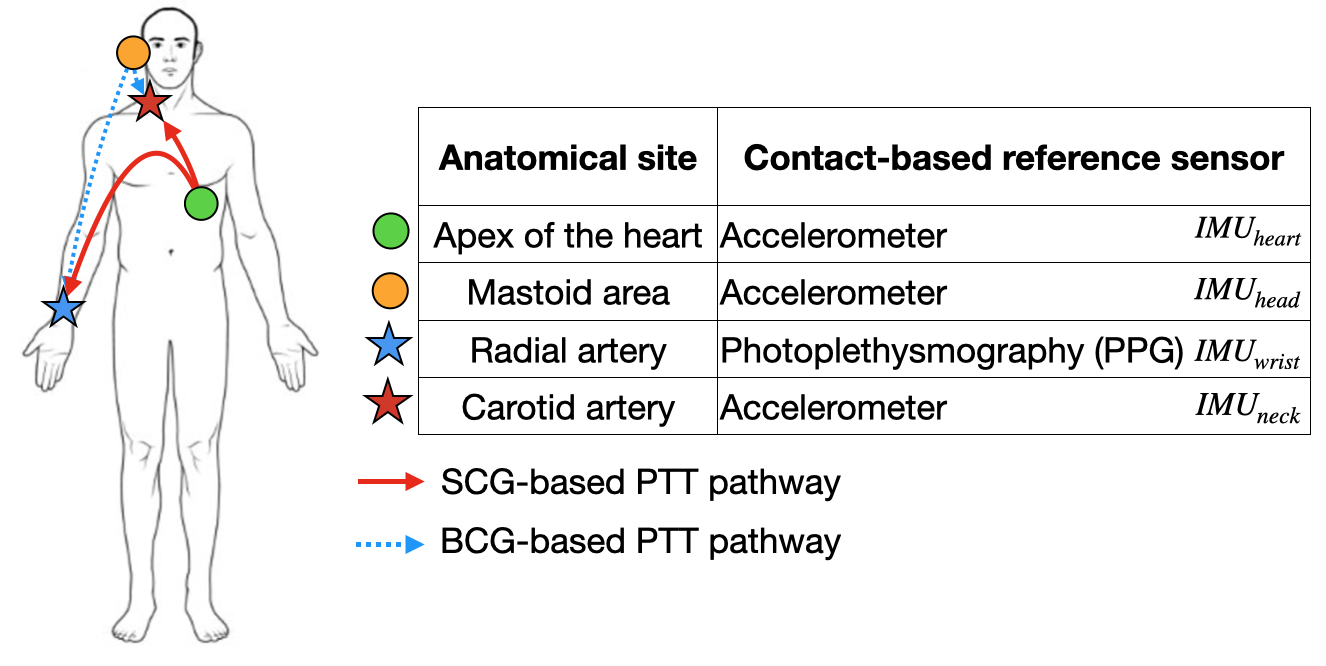}}
\caption{{\bf Placement of contact-based reference sensors at different anatomical sites.} (Image source: https://www.template.net/design-templates/print/free-body-diagram/)}
\label{fig:ref_sensors}
\end{figure}

\clearpage
\newpage

\begin{figure}[H]
\centering
{\includegraphics[width=.95\textwidth]{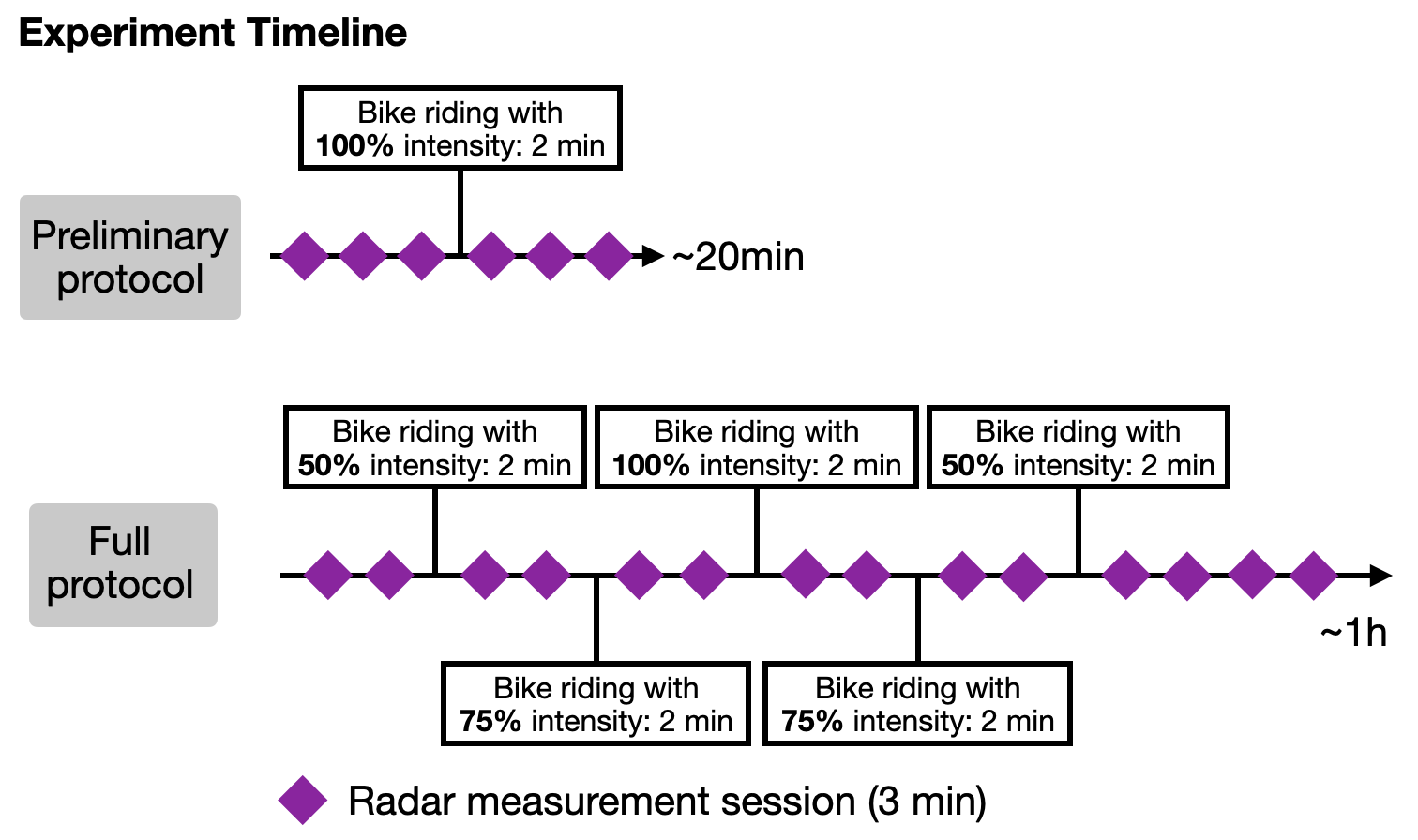}}
\caption{{\bf Experiment protocol timeline.} Ten participants took part in the preliminary protocol, while 27 took part in the full experimental protocol.}
\label{fig:timeline}
\end{figure}

\clearpage
\newpage

\begin{figure}[H]
\centering
{\includegraphics[width=.95\textwidth]{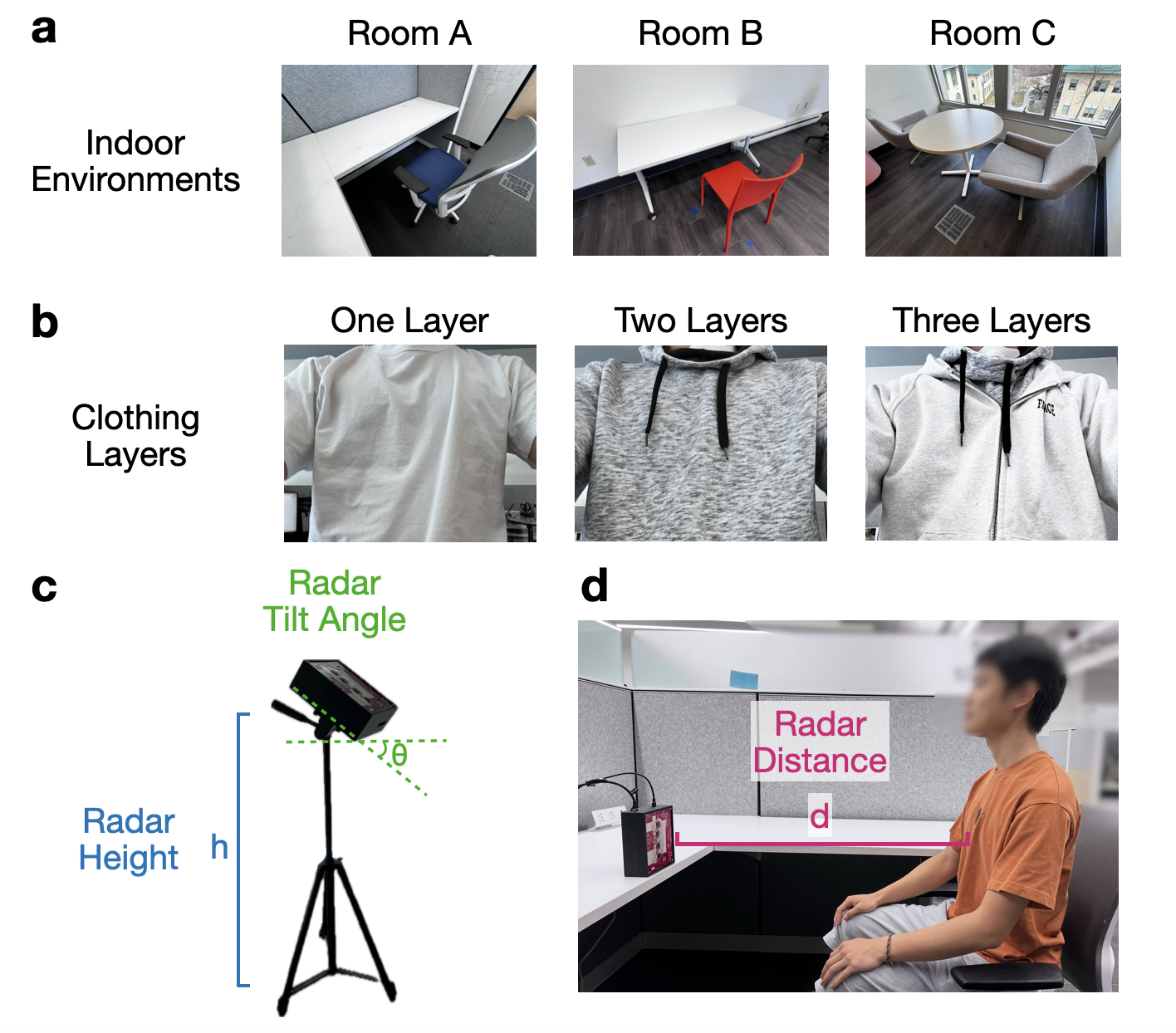}}
% \vspace{-1em}
\caption{{\bf Experimental conditions and setup for benchmark testing.}}
\label{fig:benchmark}
\end{figure}

\begin{figure}[H]
\centering
{\includegraphics[width=0.67\textwidth]{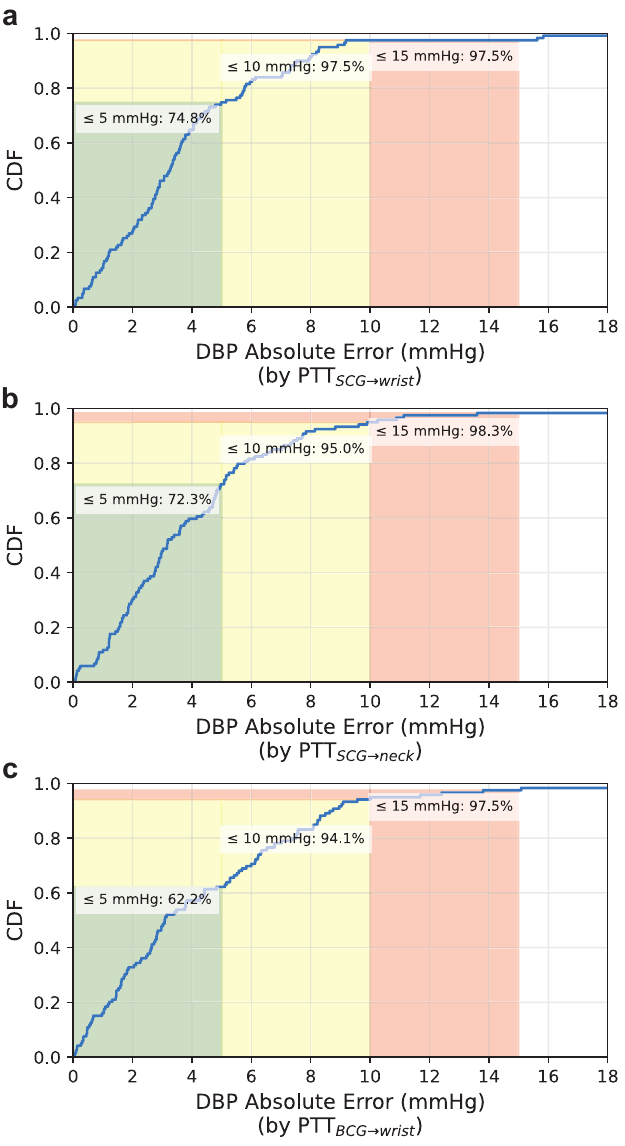}}
\caption{{\bf Cumulative distribution function of Diastolic Blood Pressure~(DBP) error.} The DBP errors estimated by three PTT pairs ($\pttscgwrist$, $\pttscgneck$, and $\pttbcgwrist$) all meet the requirements for Grade A classification by the British Hypertension Society (BHS) criteria (at least 60\% of measurements must be $\leq$ 5~mmHg, 85\% be $\leq$ 10~mmHg, and 95\% be $\leq$ 15~mmHg). }
\label{fig:bhs}
\end{figure}

\begin{figure}[H]
\centering
{\includegraphics[width=\textwidth]{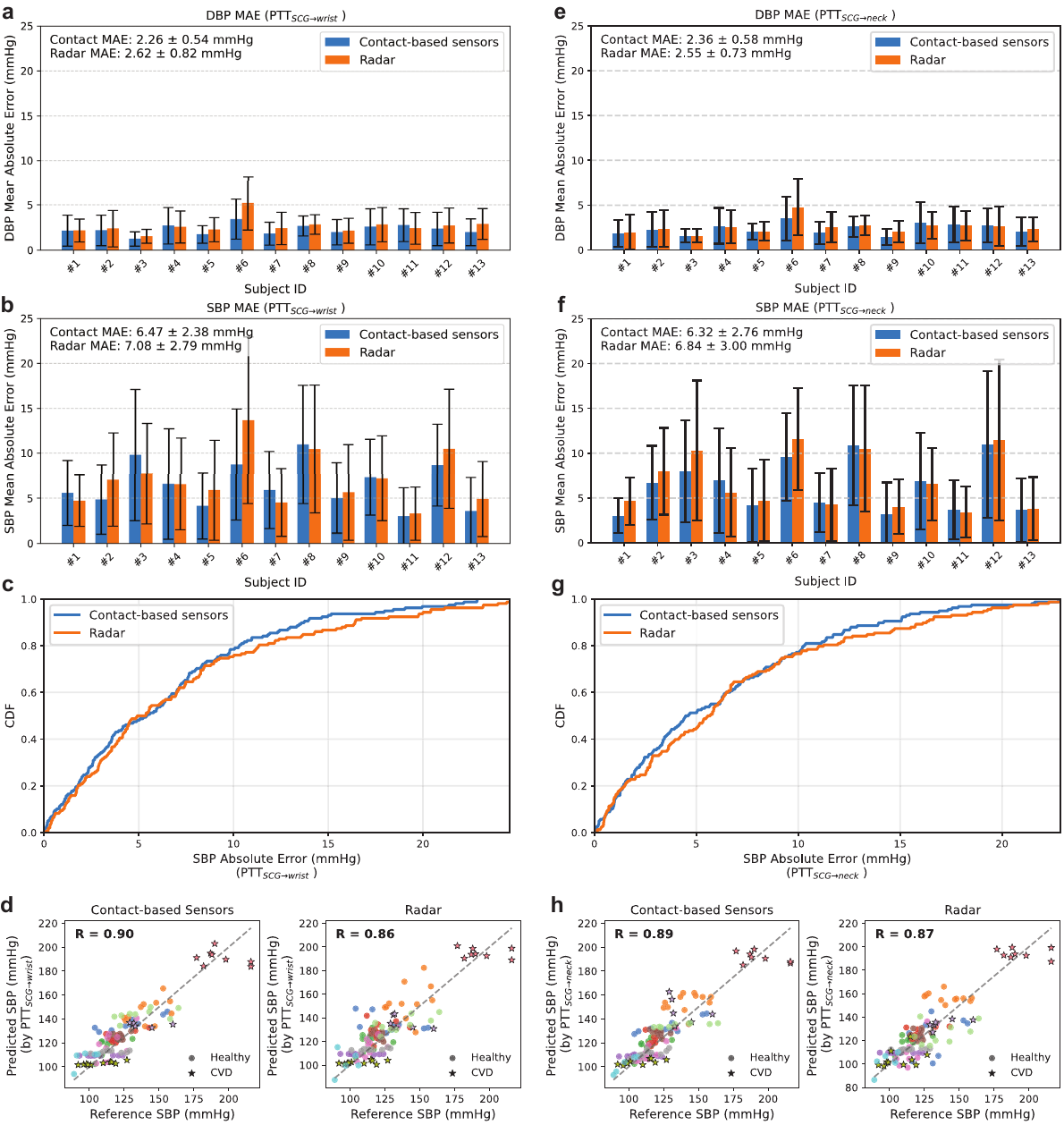}}
\caption{{\bf Systolic Blood Pressure (SBP) and Diastolic Blood Pressure (DBP) estimation errors of contact-based reference sensor and our radar systems based on best linear fit with $\pttscgwrist$~(a, b, c) and $\pttscgneck$~(e, f, g)}. {\bf a, e} shows the DBP estimation errors are consistently low across subjects, while {\bf b, f} shows the SBP estimation errors varies more across subjects, for both contact-based sensors and our radar-based system. {\bf c, g} present the cumulative distribution function of SBP error. {\bf d, h} show the SBP correlation on both contact-based sensors and radar system. The performance of our radar-based system is comparable to that of contact-based sensors.}
\label{fig:sbp_compare}
\end{figure}

\clearpage
\newpage

\begin{table}[ht]
\centering
\renewcommand{\arraystretch}{1.5}
\begin{tabular}{|l|l|}
\hline
\textbf{Parameter} & \textbf{Value} \\
\hline
\texttt{numADCSample} & $128$ \\
\hline
\texttt{startFreqConst} & $77$ GHz \\
\hline
\texttt{chirpSlope} & $63.343$ MHz/µs \\
\hline
\texttt{chirpIdleTime} & $97$ µs \\
\hline
\texttt{adcStartTimeConst} & $6$ µs \\
\hline
\texttt{chirpRampEndTime} & $63.14$ µs \\
\hline
\texttt{framePeriodicty} & $0.1$ s \\
\hline
\texttt{nchirp\_loops} & 50 \\
\hline
\end{tabular}
\caption{{\bf Radar configuration parameters.}}
\label{tab:radar_config}
\end{table}

\clearpage
\pagebreak

% \begin{table}[ht]
% \centering
% \begin{tabular}{|c|c|}
% \hline
% \textbf{Age} & 29.0 (IQR: 34.0) years \\
% \hline
% \textbf{Height} & 172.72 (IQR: 7.62) cm \\
% \hline
% \textbf{Weight} & 75.3 (IQR: 15.88) kg \\
% \hline
% \textbf{Female-to-male ratio} & 0.86 \\
% \hline
% \end{tabular}
% \caption{Demographic Summary on Subject Test Set}
% \end{table}

% \clearpage
% \pagebreak

% \begin{figure}[ht]
%     \centering
%     \includegraphics[width=\textwidth]{figs/demographicBarPlot.png}  % Adjust width as necessary
%     \caption{Demographic Bar Plot}
% \end{figure}

\setlength{\tabcolsep}{3pt} % decrease spacing between columns
\renewcommand{\arraystretch}{2.7}  % increase row height
\begin{table}[ht]
    \centering
    \footnotesize
    \caption{Demographic summary of participants in the evaluation cohort phase.}
    \begin{threeparttable}
    \label{tab:demo_summary}
    \begin{tabular}{|l||c|c|c|c|c|c|c|c|c|c|}
    \hline
    \shortstack{\bf Subject \\ (\#)} & 
\shortstack{\bf Height \\ (cm)} & 
\shortstack{\bf Weight \\ (kg)} & 
\shortstack{\bf BMI \\ (kg/m$^2$)} & 
\bf Gender & 
\shortstack{\bf Age \\ (yrs)} & 
\bf Race & 
\shortstack{\textbf{SBP range} \\ \textbf{(mmHg)}} & 
\shortstack{\textbf{DBP range} \\ \textbf{(mmHg)}} & 
\shortstack{\textbf{HR range} \\ \textbf{(bpm)}} & 
\bf CVD \\ \hline
    1  & 185 & 75  & 21.9 & Female & 24 & White   & 104--135 & 67--79   & 50--90   & -- \\ \hline
    2  & 170 & 64  & 22.2 & Female & 72 & White   & 92--131  & 53--68   & 63--72   & AF \\ \hline
    3  & 173 & 84  & 28.1 & Male   & 58 & White   & 118--174 & 70--81   & 61--97   & -- \\ \hline
    4  & 183 & 111 & 33.2 & Male   & 73 & White   & 129--160 & 84--99   & 69--80   & AF, T2D \\ \hline
    5  & 160 & 71  & 27.7 & Female & 29 & Black   & 112--143 & 74--85   & 72--105  & -- \\ \hline
    6  & 170 & 56  & 19.4 & Female & 24 & Asian   & 104--156 & 82--102  & 80--128  & -- \\ \hline
    7  & 173 & 73  & 24.4 & Male   & 31 & Asian   & 101--129 & 70--81   & 63--86   & -- \\ \hline
    8  & 188 & 100 & 28.3 & Male   & 69 & White   & 177--221 & 109--118 & 67--79   & HTN \\ \hline
    9  & 165 & 50  & 18.4 & Female & 24 & Asian   & 87--117  & 69--77   & 72--135  & -- \\ \hline
    10  & 168 & 75  & 26.6 & Female & 55 & White   & 93--125  & 78--91   & 87--117  & -- \\ \hline
    11 & 173 & 79  & 26.4 & Male   & 24 & Asian   & 115--138 & 80--91   & 80--132  & -- \\ \hline
    12 & 175 & 83  & 27.1 & Male   & 24 & Asian   & 125--181 & 79--92   & 69--129  & -- \\ \hline
    13 & 170 & 67  & 23.2 & Male   & 24 & Asian   & 104--131 & 72--87   & 67--95   & -- \\ \hline 
    \shortstack{Mean \\ (range)}
     & \shortstack{173 \\ (160--188)} 
     & \shortstack{76 \\ (50--111)} 
     & \shortstack{25.2 \\ (18.4--33.2)} 
     & -- 
     & \shortstack{41 \\ (24--73)} 
     & -- 
     & \shortstack{-- \\ (87--221)} 
     & \shortstack{-- \\ (53--118)} 
     & \shortstack{-- \\ (50--135)} 
     & --\\ \hline
    \end{tabular}
    \begin{tablenotes}
      \footnotesize
      \item BMI = body mass index, SBP = systolic blood pressure, DBP = diastolic blood pressure, HR = Heart Rate, \\ CVD = Cardiovascular Disease, AF = atrial fibrillation, T2D = Type 2 Diabetes, HTN = Hypertension
    \end{tablenotes}
  \end{threeparttable}
\end{table}

\clearpage
\pagebreak

% \begin{algorithm}[H]\label{algo:cross_correlation}
% \caption{Algorithm to time-align sensor data from accelerometers pairs $IMU_{head}$, $IMU_{chest}$, and $IMU_{chest}$, $IMU_{wrist}$, as well as $IMU_{wrist}$ and the mmWave radar data.}
% \DontPrintSemicolon
% \SetKwInOut{Input}{Input}\SetKwInOut{Output}{Output}
% \SetKwFunction{FMain}{time\_align}
% \SetKwProg{Fn}{Function}{:}{}
% \Fn{\FMain{data\_1,data\_2}} {

% \begin{algorithmic}
% \State
% \State \Input{$data\_1$ Sensor data}
% \State \Input{$data\_2$ Sensor data}

% \State $w\_len=2$~s\;
% \State $step=0.02$~s\;
% \State $L=length(data\_1)$\;
% \State $r\_vals=[]$\;
% \State $lag\_vals=[]$\;

% \State \For{$w = [0:step:L-step]$} {\\
% \State $window\_1 = data\_1[w:w+w\_{len}]$\;
% \State $window\_2 = data\_2[w:w+w\_{len}]$\;
% \State $[r,lag] = cross\_corr(window\_1,window\_2)$\;
% \State $r\_{vals}.append(max(r))$\;
% \State $lag\_{vals}.append(lag[argmax(r)])$\;
% }\\
% \State $best\_lag = lag\_vals[argmax(r\_vals)]$\;

% \State \If{$best\_lag > 0$} {
%     \State $aligned\_data\_1 = data\_1[:-best\_lag]$\;
%     \State $aligned\_data\_2 = data\_2[best\_lag:]$
% }
% \State \Else {
%     \State $aligned\_data\_1 = data\_1[-best\_lag:]$\;
%     \State $aligned\_data\_2 = data\_2[:best\_lag]$
% }

% \State \Return $aligned\_data\_1, aligned\_data\_2, best\_lag$

% \end{algorithmic}
% }
% \end{algorithm}

% \clearpage
% \pagebreak
% \section*{Pseudocode for Target Bin Identification Algorithm}

\begin{algorithm}
\caption{Cardiac signal conditioning}
\label{alg:signal_processing_1}

\KwIn{
$\mathbf{s}_{r,\theta}$: Raw signal data at range $r$ and angle $\theta$ \\
$f_s$: Signal sampling frequency
}
\KwOut{
$\mathbf{s}_{\text{processed}}$: Signal after filtering and processing
}

\textbf{Phase Computation:}\\
$\theta_{\text{bin}} \gets \text{angle}(\mathbf{s}_{r,\theta})$\\
$\theta_{\text{bin}} \gets \text{unwrap}(\theta_{\text{bin}})$

\textbf{Filtering:}\\
Apply Butterworth band-pass filter with cutoff frequencies $f_c = \text{filter\_range}$ and a 1st-degree polynomial

\textbf{Detrending:}\\
$\mathbf{s}_{\text{processed}} \gets \text{detrend}(\mathbf{s}_{r,\theta})$ using a 4th-degree polynomial

\Return $\mathbf{s}_{\text{processed}}$

\end{algorithm}

\begin{algorithm}
\caption{Identification of candidate cardiac signal regions}
\label{alg:find_bin}

\KwIn{
$\mathbf{S}$: Signal Bins (time $\times$ range $\times$ angle)\\
$f_s$: Signal Sampling Frequency\\
$R_{\text{target}}$: Target Range\\
$\Theta_{\text{target}}$: Target Angle\\
$f_{\text{BPM}}$: Target Heart Beat Frequency\\
$T$: Target Points (Heart, Head, Neck, Wrist)
}
\KwOut{
$\mathbf{H}$: History of Detected Bin Parameters
}

\textbf{Initialization:} Set $\mathbf{H}$ to an empty list

\textbf{Range-Angle Bin Search:}\\
\For{$r \in \text{ValidRangeBins}(R_{\text{target}})$}{
    \For{$\theta \in \text{ValidAngleBins}(\Theta_{\text{target}})$}{
        Extract Bin Signal: $\mathbf{s}_{r,\theta}$\\
        $M_{r,\theta} \gets \text{Magnitude}(\mathbf{s}_{r,\theta})$\\
        $\Phi_{r,\theta} \gets \text{Phase}(\mathbf{s}_{r,\theta})$\\
        $\mathbf{s}_{\text{processed}} \gets \text{SignalPreprocessing}(\mathbf{s}_{r,\theta}, f_s)$

        \If{$T_\text{Heart} \lor T_\text{Neck}$}{
            $\mathbf{s}_{\text{processed}} \gets \text{DerivativeFilter}(\mathbf{s}_{\text{processed}})$
        }

        $[\mathbf{ACF}, \mathbf{Lag}] \gets \text{Autocorrelation}(\mathbf{s}_{\text{processed}})$\\
        $\mathbf{P} \gets \text{findpeaks}(\mathbf{ACF}, f_{\text{BPM}})$\\
        $[P_{\text{max}}, \text{index}_{\text{max}}] \gets \max(\mathbf{P})$\\
        $Lag_{\text{max}} \gets \mathbf{Lag}[\text{index}_{\text{max}}]$\\
        Record Results in $\mathbf{H}$: $[P_{\text{max}}, Lag_{\text{max}}, M_{r,\theta}, r, \theta]$
    }
}
Sort $\mathbf{H}$ by $P_{\text{max}}$ (Descending Order)\\
\Return $\mathbf{H}$

\end{algorithm}

\begin{algorithm}
\caption{Identification optimal cardiac signal region}
\label{alg:signal_processing_2}

\KwIn{
$\mathbf{H}$: History of Detected Bin Parameters \\
$\mathbf{T}$: Target Points (Heart, Head, Neck, Wrist)
}
\KwOut{
$\mathbf{H_{selected}}$: Selected list with corresponding bin info
}

\If{$T_\text{Heart} \lor T_\text{Neck}$}{
    $\mathbf{H_{selected}} \gets \text{Select top k bins from } \mathbf{H}$\\
    $\mathbf{H_{selected}} \gets \text{Sort rows of } \mathbf{H} \text{ by the 3rd column (Magnitude) in descending order}$\\
    $\mathbf{H_{selected}} \gets \text{Select top } N \text{ rows from the sorted list}$
}

\If{$T_\text{Wrist}$}{
    $\mathbf{H_{selected}} \gets \text{Sort rows of } \mathbf{H} \text{ by the 2nd column } (\text{Lag}_{\text{max}}) \text{ in descending order}$
    $\text{most\_frequent\_lags} \gets \text{most\_appear\_number}(\mathbf{H})$\\
    $\mathbf{H_{selected}} \gets \text{Select rows of } \mathbf{H} \text{ where the 5th column (Peak Lags) matches } \text{most\_frequent\_lags}$
}

\If{$T_\text{Head}$}{
    $\mathbf{H_{selected}} \gets \text{Select top k bins from } \mathbf{H}$\\
    $r_{mid} \gets mid\{\,r_i  \in history\_list\}$\\
    $\theta_{mid} \gets mid\{\theta_i \in history\_list\}$\\
    $\mathbf{H_{selected}} \gets \text{Select rows of } \mathbf{H} \text{ where }r_i \text{ or } \theta_i  \text{ is close to } r_{mid} \text{ or } \theta_{mid}$
}

\Return $\mathbf{H_{selected}}$

\end{algorithm}

\end{document}